\documentclass[journal, final]{IEEEtran}
\usepackage{graphicx}
\usepackage{soul}
\usepackage{color}
\usepackage{amsmath}
\usepackage{comment}
\usepackage{cite}
\usepackage{multirow}
\usepackage{bm}
\usepackage[acronym,shortcuts]{glossaries}
\usepackage[linesnumbered,lined, algoruled]{algorithm2e}
\usepackage{subfigure}
\newacronym{acn}{ACN}{Adaptive Charging Network}
\newacronym{cps}{CPS}{Cyber-Physical System}
\newacronym{cts}{CTS}{Current Time Series}
\newacronym{dt}{DT}{Decision Tree}
\newacronym{ev}{EV}{Electric Vehicle}
\newacronym{evse}{EVSE}{Electric Vehicle Supply Equipment}
\newacronym{knn}{kNN}{k-Nearest Neighbors}
\newacronym{rf}{RF}{Random Forest}
\newacronym{rn}{ResNet}{Resolution Network}
\newacronym{rsse}{RSSE}{residual sum of squared errors}
\newacronym{soc}{SoC}{State of Charge}
\newacronym{svm}{SVM}{Support Vector Machine}
\newacronym{ts}{TS}{time series}
\newacronym{ada}{ADA}{ADA Boost}
\newacronym{lr}{LR}{Logistic Regression}
\setacronymstyle{short-long}

\title{\emph{EVScout2.0}: Electric Vehicle Profiling Through Charging Profile}

\author{Alessandro Brighente,Mauro Conti,Denis Donadel,Federico Turrin \thanks{Federico Turrin is supported by a grant from the Cariparo Foundation and Yarix S.r.l. which we would like to thank.} \\
Department of Mathematics, University of Padova, Padova, Italy}

\begin{document}
\maketitle
\begin{abstract}

\acp{ev} represent a green alternative to traditional fuel-powered vehicles. To enforce their widespread use, both the technical development and the security of users shall be guaranteed. Privacy of users represents one of the possible threats impairing \acp{ev} adoption. In particular, recent works showed the feasibility of identifying \acp{ev} based on the current exchanged during the charging phase. In fact, while the resource negotiation phase runs over secure communication protocols, the signal exchanged during the actual charging contains features peculiar to each \ac{ev}. A suitable feature extractor can hence associate such features to each \ac{ev}, in what is commonly known as profiling.

In this paper, we propose \emph{EVScout2.0}, an extended and improved version of our previously proposed framework to profile \acp{ev} based on their charging behavior. By exploiting the current and pilot signals exchanged during the charging phase, our scheme is able to extract features peculiar for each \ac{ev}, allowing hence for their profiling. We implemented and tested \emph{EVScout2.0} over a set of real-world measurements considering over 7500 charging sessions from a total of 137 \acp{ev}. In particular, numerical results show the superiority of \emph{EVScout2.0} with respect to the previous version. \emph{EVScout2.0} can profile \acp{ev}, attaining a maximum of $0.88$ recall and $0.88$ precision. To the best of the authors' knowledge, these results set a new benchmark for upcoming privacy research for large datasets of \acp{ev}.

 \end{abstract}


\glsresetall

\section{Introduction}
\acp{ev} represent one of the technologies enabling a long-term solution for the mitigation of petroleum consumption and the consequent reduced emissions amount. Multiple countries already started providing financial incentives to facilitate the purchasing and widespread of \acp{ev} ownership~\cite{wang2019global}. The \ac{ev} global forecast expects a compound annual growth rate of $29\%$ over the next ten years, with a total \ac{ev} sales reaching up to $1.1$ million by $2030$~\cite{deloitte}. Among the factors abating the adoption of \acp{ev}, consumers' concerns regard the availability of charging infrastructures~\cite{deloitte}. In fact, while gas stations are widely deployed both in cities and rural areas, \acp{evse} are not deployed in a sufficient number or, in some areas, are not present at all. Although charging stations may be deployed at house premises, the absence of publicly available \acp{evse} represent a limit, as a user is forced to limit the traveling distance to stay close to a charging point. To mitigate this issue and increase the interest towards \acp{ev}, current recovery plans envisioned by countries such as Germany and China designate part of their funds to the development of \ac{ev} charging infrastructures~\cite{deloitte}. Furthermore, companies are equipping their parking spots with \acp{evse} to serve employees \acp{ev}. For instance, United States is incentivizing the adoption of \acp{evse} at companies' parking spots via the workplace charging challenge~\cite{evChall}.  Therefore, in the next years, we expect a significant increase in the number of publicly available \acp{evse}, allowing users to charge their \ac{ev} at any time, removing availability concerns.

\acp{evse} enable the charging process by bringing together multiple technologies. On the one hand, they allow the user to exchange data with the grid, providing means for authorizations towards a central entity to negotiate the service and pay the associated fees. On the other hand, they allow for an exchange of information from the \ac{ev} to the infrastructure, such that the charging process is conducted by providing safety towards \ac{ev}'s components. The former communication process (i.e., user to infrastructure) is secured by means of cryptographic procedures as well as secure network protocols. Their use mitigates all the very well-known threats towards the users' security and privacy given by unprotected communications. However, the latter communication process (i.e., \ac{ev} to \ac{evse}) is not secure, as signals are exchanged without encryption or aggregation techniques. The signals exchanged during the charging process can hence be exploited as a side-channel to extract information peculiar to each \ac{ev}, allowing hence for their profiling and successive recognition~\cite{brighente2021}. This represents a threat towards users' privacy, as the connection of their \ac{ev} to an \ac{evse} monitored by a malicious user may lead to tracking their movement as well as information regarding their driving behavior. Since the majority of publicly available \acp{evse} are deployed without proper physical protection, they can be accessed by anyone and represent hence favorable spots for attackers targeting the charging infrastructure~\cite{antoun2020detailed}. Therefore, an attacker can easily install devices to collect data regarding the charging process. 

In this paper, we propose \emph{EVScout2.0}, a framework extending EVScout, the work proposed in~\cite{brighente2021}, where we initially showed the feasibility of profiling \acp{ev} based on the current exchanged during the charging process. In particular, we extend the proposed framework by developing an enhanced feature extractor, which allows for higher classification scores compared to our previous work. We then exploit a novel \ac{ts} for feature extraction, given by the combination of the current and pilots \acp{ts}. We will explain the reasoning behind this choice and provide the details on its computation. Furthermore, we consider a larger real-world dataset, comprising up to $300$ \acp{ts} of the current and pilot signals exchanged by each of the $137$ considered \acp{ev}, for a total of more than $7500$ charging session. We perform a thorough evaluation of \emph{EVScout2.0}, showing its profiling performance considering different training set sizes, as well as different unbalancing in the training-testing datasets. Compared to the work in~\cite{brighente2021}, we provide a more comprehensive analysis of the performance of the attack. In particular, we first show the performance of the novel feature extractor and investigate the performance of \emph{EVScout2.0} for a varying number of features. We then compare the performance of the different classifiers that can be exploited by \emph{EVScout2.0}. Compared to \cite{brighente2021} we also extend the number of classifiers and provide an in-depth description of the choice of their hyper-parameters. We then investigate the dependencies of \emph{EVScout2.0} on the number of training \acp{ts} needed for classifying \acp{ev} with sufficient confidence. We show that the attack is already successful considering $7$ training examples. We then investigate the battery degradation over time and its impact on \emph{EVScout 2.0} performances. Lastly, we show the superiority of \emph{EVScout 2.0}, comparing its performance with those in~\cite{brighente2021}, both on the old and new datasets.

The contribution of this paper and its improvements with respect to~\cite{brighente2021} can hence be summarized as follows:
\begin{itemize}
    \item We propose an enhanced feature extraction framework, which allows for better classification performance compared to~\cite{brighente2021}.
    \item We propose the use of Delta \ac{ts}, a novel \ac{ts} given by the linear combination of the current and pilot \acp{ts}. Delta \ac{ts} allows for the extraction of more significant features compared to those in \cite{brighente2021}.
    \item We analyze the performance of \emph{EVScout2.0} on a large real-world dataset, comprising more than $7500$ employable current and pilot time series from $137$ \acp{ev}. Compared to the dataset in~\cite{brighente2021}, we consider both a larger number of \acp{ts}, as well as a larger number of \acp{ev}.
    \item We perform a thorough evaluation of \emph{EVScout2.0}, showing its performance over different training set sizes, as well as its robustness towards different unbalancing amounts in the training-testing dataset sizes.
    \item Compared to~\cite{brighente2021}, we analyze the performance of a higher number of classifiers and provide an in depth discussion on the choice of their hyper-parameters. We show that \emph{EVScout2.0} is able to profile \acp{ev} with precision and recall up to $0.88$.
    \item We investigate the battery degradation over time, and show that \emph{EVScout2.0} is able to extract features that allow for high classification scores (on average $0.8$ $F1$ score) in time.
    \item We compare \emph{EVScout2.0} with the framework in~\cite{brighente2021}, showing its superiority both in the old and in the new datasets.
\end{itemize}

The rest of the paper is organized as follows. In Section~\ref{sec:related} we review the related works. Then, in Section~\ref{sec:sys_threat} we present the considered system and threat model in an \ac{ev} charging infrastructure. In Section~\ref{sec:attack} we describe the features and the steps of \emph{EVScout2.0}. In Section~\ref{sec:evaluation} we describe the evaluation framework and the new dataset. Then, in Section~\ref{sec:profiling} we show the results achieved by \emph{EVScout2.0}, while in Section~\ref{sec:additional_analysis} we proposed some insight on additional analysis related to training set size and battery degradation.
In Section~\ref{sec:comparison} we provide a comparison between \emph{EVScout2.0} and its previous version EVScout~\cite{brighente2021}, clearly presenting the advantages of the new approach. We discuss some countermeasures to avoid profiling in Section~\ref{sec:countermeasures} and, lastly, we sum up the results and draw the conclusions in Section~\ref{sec:conclusions}. 

\section{Related Works}\label{sec:related}

Power consumption can be exploited as a side-channel for different purposes~\cite{LIU2021100007}. For instance, an attacker may implement a laptop user recognition by exploiting the current drawn by a smart wall socket during users' activity given~\cite{conti2016mind}. The same concept can be exploited for detecting the presence of the user in a smart home, where raw data can be acquired and analyzed to detect for activity and hence users' presence~\cite{molina2010private}. Raw power data also provides information regarding the actions a user is performing. For instance, by analyzing raw power data exchanged via a USB cable, an attacker may be able to infer the activity of a user on a smartphone~\cite{spolaor2017no} or to obtain information regarding the victim's browsing activities~\cite{yang2016inferring}. The power exchanged during the charging process via USB can also leak more sensitive information, which an attacker can later exploit. For instance, the power analysis may leak information regarding digits composed on a touchscreen, allowing for the deduction of users' passwords~\cite{cronin2021charger}.

Regarding \acp{evse}, security and privacy research and contributions focus on the negotiation phase, as it allows the negotiation of the charging service by sharing personal user's data. This includes both communications from the \ac{evse} to the power distributor and from the \ac{ev} to the \ac{evse}. In the former communication scenario, the standards J1772~\cite{std2001j1772} and  CHAdeMO~\cite{anegawa2010characteristics} are exploited to regulate the physical standards needed for \ac{ev} to \ac{evse} connections, together with the signaling required for the charging process. These standards do not employ encryption or privacy measures on the exchanged information. In the latter case, the standard ISO 15118~\cite{multin2018iso} is exploited to create a secure communication link, implying that both \ac{ev} and \ac{evse} must be able to encrypt messages. The overall system has been analyzed in literature from a \ac{cps} security point of view, and several threats have been identified~\cite{antoun2020detailed, gottumukkala2019cyber}. However, security and privacy analysis should also focus on the charging phase. In fact, the signals exchanged during the charging process create a time series that can be analyzed to extract features that lead to the profiling of vehicles~\cite{brighente2021}. However, an in-depth analysis of the robustness of these types of attacks is still missing. In particular, the feasibility of the attacks has been shown for a limited dataset, and the requirements of the attacker in terms of the number of samples needed for \ac{ev} profiling has not been discussed yet.

\section{System and Threat Model}\label{sec:sys_threat}

In this section we introduce the scenario on which we conceived our experiments. In particular, in Section~\ref{subsec:system_model} we recall the \ac{ev} charging system, which represents our system model, then in Section~\ref{subsec:threat_model} we present the threat model designed for \emph{EVScout2.0}.

\subsection{EV Charging System}\label{subsec:system_model}

\begin{figure}[h]
    \centering
    \includegraphics[width=0.8\columnwidth]{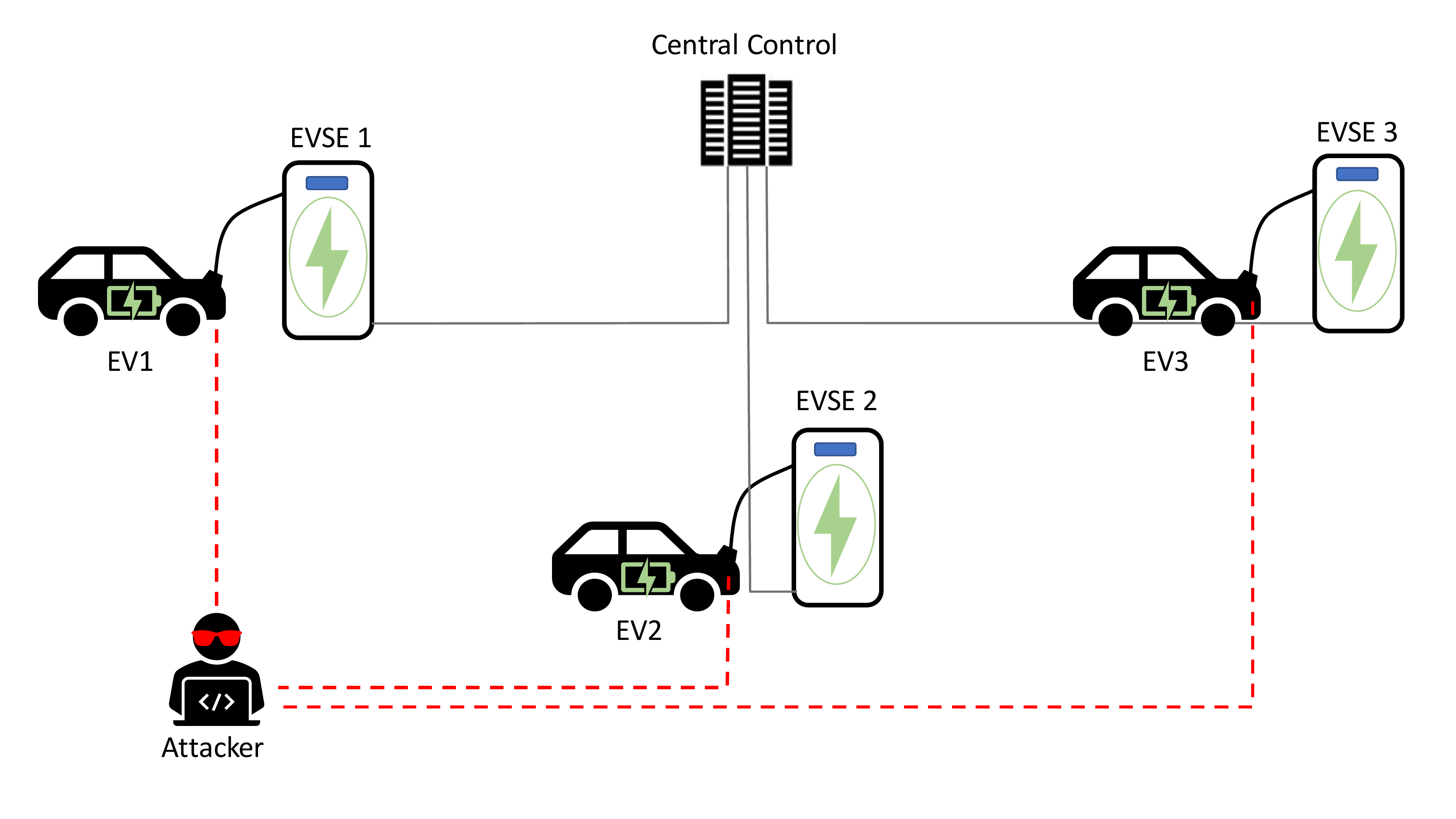}
    \caption{System and threat model. Multiple \acp{evse} are connected to the central control which provides coordination and power distribution among them. A single \ac{ev} is connected to each \ac{evse}. The attacker has access to the physical quantities exchanged by multiple \acp{ev} during the charging phase.}
    \label{fig:thModel}
\end{figure}

According to the Vehicle-To-Grid (V2G) paradigm~\cite{Kempton2001}, the charging infrastructure for \acp{ev} is a network where a central controller (power distributor) distributes power based on \acp{evse} demand while accounting for the maximum supported load by the electric grid. We depicted in Figure~\ref{fig:thModel} the typical architecture of a V2G system.
\acp{evse} in the network may be deployed at different sites, e.g., private customer premises, public stations, or office buildings. Each \ac{ev} is both physically and logically connected to the grid via the \ac{evse}, which manages communications between the user (i.e., the owner of the \ac{ev}) and the power distributor. For public charging infrastructures and office stations, multiple \acp{evse} are connected to the power distributor through a Central Control that copes with the demand of a large number of connected users~\cite{antoun2020detailed}. \acp{evse} are typically equipped with communication interfaces (wireless or wired) to allow communication with the user and the grid. Utilizing modules in the \ac{ev} or smartphone, the user can communicate with the \ac{evse} and, in turn, with the power supplier.

Current implementations of \acp{evse} are organized in three levels~\cite{gottumukkala2019cyber,wang2021grid}. Level $1$ and $2$ use a $5$ lead connector based on SAE J$1772$ standard~\cite{Troepfer2009}, where $3$ leads are connected to the grid via relays in the \ac{evse}. The remaining $2$ pins, i.e., pilot and proximity lines, are used for signaling. The proximity line indicates whether a good physical connection has been established between the \ac{ev} and the \ac{evse}, blocking the initiation of the charging process in case devices are not properly attached and preventing hence damages to both the user and the involved devices. The pilot line provides a basic communication means between the \ac{ev} and the \ac{evse}.
The combination of signals collected from all the pins is used to provide the main processing unit of the \ac{evse} information regarding the charging process, allowing for metering used to assess the charging session state.
If a problem arises at one of the two sides of the charging process, the \ac{evse} computer hardware will remove power from the adapter to prevent injuries on both sides. Level $3$ \acp{evse} are instead more complex, comprising bigger pins for power delivery and allowing power line communications via the pilot line.

\begin{figure}[t]
    \centering
    \includegraphics[width = 0.6\columnwidth]{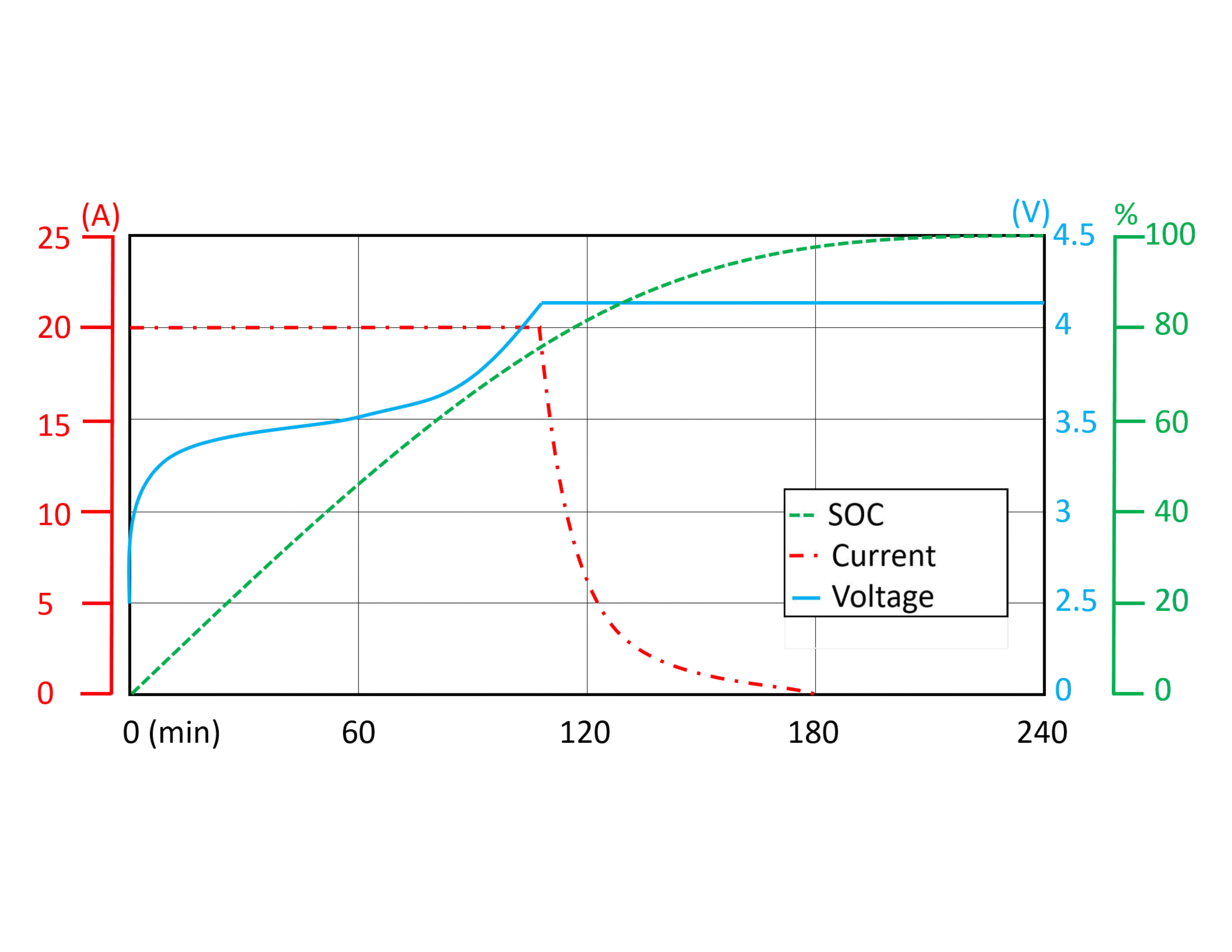}
    \caption{Charging profile of a Li-ion battery~\cite{thundersky}: we see that, as the \ac{soc} increases, the charging mode switches from constant current to constant voltage. We further notice that the two phases are mutually exclusive.}
    \label{fig:soc}
\end{figure}

Typical batteries employed for \acp{ev} belong to the class of Li-ion (lithium-ion)~\cite{bai2014experiments,wu2017optimization}. Current and voltage values exchanged during the charging process depend on the \ac{soc} of the \ac{ev} battery and can be divided into two classes: constant current/constant voltage and constant power/constant voltage~\cite{marra2012demand}. In this work, we consider the first class, where the charging process can be further divided into two phases: 
\begin{itemize}
    \item \textit{Constant current phase}, where the current level is constant while the voltage value increases;
    \item \textit{Constant voltage phase}, where voltage is constant whereas current decreases.
\end{itemize}
The charging process starts with the constant current phase, and this operation mode is kept until the battery's \ac{soc} is below a certain value. After reaching the \ac{soc} switching point, the operation mode switches to constant voltage up to the full charge. Typical \ac{soc} switching values lie between $60\%$ and $80\%$ of the full charge. An example of a charging profile for an \ac{ev}'s Li-ion battery is shown in Figure~\ref{fig:soc}. We here remark that constant current and constant voltage phases are mutually exclusive in time, as this will be exploited by \emph{EVScout2.0}. 

\subsection{Threat Model}\label{subsec:threat_model}

We assume a threat model similar to ATM skimming, where the attacker is equipped with a small measuring device that can be connected on one side to the \ac{evse} plug and the other side to the \ac{ev} plug. This device is used to measure the exchanged current at the connection point between the \ac{ev} and \ac{evse}, and we assume that it is hard to be noticed by users. Moreover, we assume that the device provides information to the attacker via either i) a wireless communication module or ii) storing the values of interest to be lately collected by the attacker. The attacker has hence access to the \ac{ts} of the signals exchanged between the \ac{evse} and the \ac{ev} during the charging phase. These values are hence recorded for each pin of the \ac{ev} charger. In this paper, we assume that the attacker trains a different classifier for each target \ac{ev}. Therefore, in order to properly train the classifier, a sufficient number of \acp{ts} shall be collected for every single \ac{ev}. The attack is hence divided into i) collection phase, where the attacker collects data regarding a target \ac{ev}, and ii) exploitation phase, where the attacker exploits the previously computed features to discriminate between different \acp{ev} based on the observed time series. To speed up the collection process, the attacker plugs a collector device in each \ac{evse}, such that signals can be simultaneously recorded from multiple sessions. The target \ac{ev} may be plugged to different \acp{evse} in successive charging sessions, e.g., in the parking spaces at workplaces or shopping malls the victim often visits. Therefore, by deploying multiple collecting devices at different \acp{evse}, the attacker has access to multiple charging sessions of the same \ac{ev} and exploits them to build a training set for feature characterization and extraction. Due to the deployment of multiple measuring devices at different \acp{evse}, the collected data is generated from the charging sessions of multiple \acp{ev}. Therefore, the attacker is able to simultaneously build a training set for multiple \ac{ev}. 

Our threat model is based on the fact that the majority of publicly available \acp{evse} are deployed without proper physical security and hence can be accessed by any malicious actor~\cite{antoun2020detailed}. In this case, since there is no access regulation to the \acp{evse}, the attacker can freely attach the measuring devices. The attack is further facilitated by the fact that typical users are not going to modify the charging system, even though they may notice something unfamiliar. Therefore, the attacker not only may be represented by the company running the \acp{evse} network, but can also be anyone interested in obtaining information on users' consumes and locations. On the other hand, we notice that the \ac{evse} devices can be routinely checked by the staff of the running company. Therefore, we argue that the most successful attacks may involve collusion in two different scenarios: 
\begin{itemize}
    \item Collusion with the running company; the attacker agrees with the staff of the running company on a certain reward they may obtain, not reporting the problem to the central office and not detaching the measuring devices.
    \item Collusion with the \ac{evse} builders; the attacker agrees on a reward to the \ac{evse} builders if measuring devices are natively built-in in the \ac{evse} plug. This guarantees that the users or the staff of the running company are unable to notice and detach the measuring device.
\end{itemize}

To build a set with sufficient features, we assume that the attacker collects the \ac{ts} of the exchanged current values and the \ac{ts} of the pilot signals. Notice that, in order to retrieve this data, the attacker does not need to perform elaborate network intrusion schemes, as signals are exchanged outside the network. Furthermore, notice that the attacker is not modifying in any way the charging process. Hence, the system cannot automatically detect the attacker's presence via intrusion/anomaly detection techniques. 

Figure~\ref{fig:thModel} shows the assumed system and threat model. In detail, multiple \acp{evse} communicate with the Central Control, which provides coordination information as well as power distribution. A single \ac{ev} is connected to each \ac{evse}. As previously mentioned, the attacker gets access to the time series of the physical quantities exchanged by multiple \acp{ev} during the charging phase and exploits them for profiling.
Noticing that if the attacker can remotely access the current exchanged in different network nodes, it can also locate users, we see how profiling may also lead to user tracking.
The knowledge of the physical signal features associated with each \ac{ev} (and hence the owning user) can also be exploited for impersonation attacks. Considering \acp{evse} which are automated based on the specific user needs, an attacker could steal assets from a target user by generating a signal with the same physical features such that the \ac{evse} recognizes the attacker as the victim. Scenarios in which this may harm the target user include billing and misbehaving users exclusion from the system. Therefore, the motivation behind the attack can be multiple. As an illustrative example, consider advertising: the attacker has both information on a certain user's typical movements and the amount of energy he/she consumes regularly. This information can be sold to \ac{evse} owners, which will target their advertisement to the profiled user according to its demand. Notice that, although a single classifier is trained for each \ac{ev}, the attacker collects information regarding multiple \acp{ev}, such that more than a single classifier can be implemented with the gathered data. Therefore, the attacker can also sell information about collective use of the \ac{evse} charging stations by \acp{ev} to \ac{evse} companies. Although profiling can be implemented using cameras, this would not allow for collecting energy traces, therefore losing some of the information available with the proposed attack. Such information can be obtained utilizing \emph{EVScout2.0}. The possibility of tracking a user gives a further threat. In fact, thanks to \emph{EVScout2.0}, a malicious user can detect the presence of a target user in a certain place and time based on the fact that her \ac{ev} is connected to a particular \ac{evse}. Although an attacker may implement this attack through cameras, the envisioned device used to gather the data needed for \emph{EVScout2.0} is less noticeable and less detectable by the running company's staff.

\section{\emph{EVScout2.0}}\label{sec:attack}

In this section we describe the \emph{EVScout2.0} analysis methodology. We propose a high-level description of the attack configuration in Section~\ref{sec:attackOverview}. Then we describe the preprocessing we apply to the dataset. Starting from Section~\ref{sec:tailsIdent}, we present the concept of tails and outline the method we designed to extract them automatically. To improve the performance with respect to the solution in \cite{brighente2021}, we propose to exploit Delta \ac{ts}, i.e., the \ac{ts} given by the combination of the current and pilot \acp{ts}. In Section~\ref{sec:delta} we present Delta \ac{ts}, providing both the motivation behind our choice and the means to compute it. Lastly, in Section~\ref{sec:featExt}, we describe the novel and automatic feature extraction technique we employ in EVScout2.0. 

\subsection{Attack Description}\label{sec:attackOverview}

As previously stated, in the context of \acp{ev} charging infrastructures, users' data is authenticated and secured. However, physical signals are generally not supposed to implement security measures and therefore can easily be exploited by malicious users. Since the exchanged current during the charging phase is a user's generated data, it comprises features and recurrent behaviors that can be exploited for profiling attacks. \emph{EVScout2.0} identifies and extracts those physical features which are representative of every single \ac{ev}, such that we can assert with sufficient confidence if and when a specific user is connected to the charging grid. 

\begin{figure}[h]
    \centering
    \includegraphics[width=0.8\columnwidth]{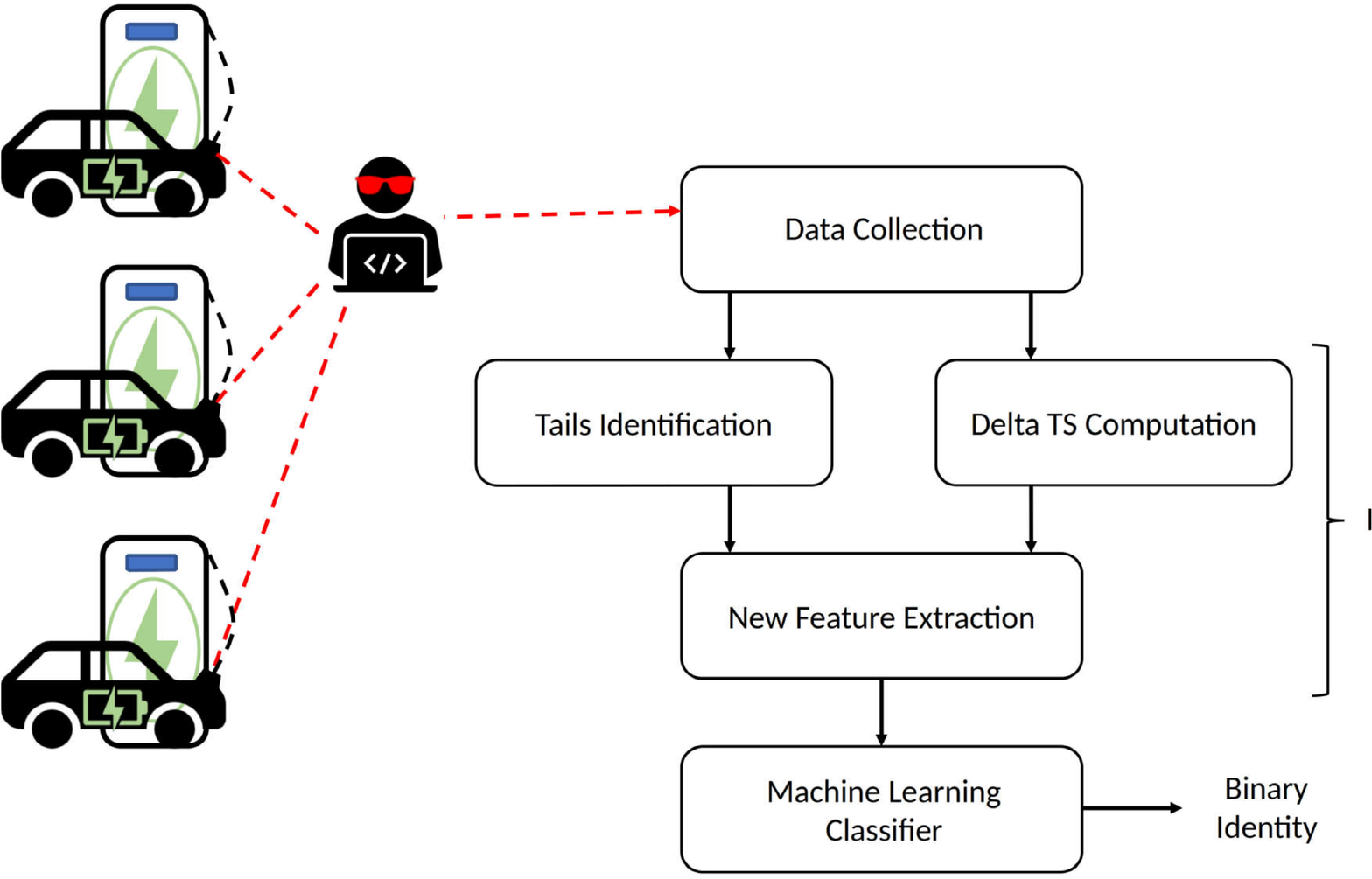}
    \caption{Block diagram of \emph{EVScout2.0}'s steps.}
    \label{fig:blocks}
\end{figure}

Figure~\ref{fig:blocks} shows the block diagram of \emph{EVScout2.0}'s steps. \emph{EVScout2.0} starts with data collection. To profile \acp{ev} the attacker must collect multiple charging sessions for each target \ac{ev}. We will discuss this requirement in Section~\ref{subsec:training_variation}, assessing the number of training examples the attacker needs to collect to profile an \ac{ev} with sufficient confidence. Once collected the charging \acp{ts} (i.e., the dataset), \emph{EVScout2.0} automatically computes the features that characterize each \ac{ev}. To this aim, in the following, we propose a strategy to exploit the behavior of the batteries during the charging process. In particular, as proposed in EVScout~\cite{brighente2021}, assuming that the attacker has access only to the ampere-based electrical quantities, we exploit the current behavior during the constant voltage phase. Leveraging the nomenclature in~\cite{sun2020classification}, we name the current \ac{ts} during the constant voltage phase as tail. In Section~\ref{sec:tailsIdent} we describe how \emph{EVScout2.0} extracts the tails.

Notice that the choice of exploiting tails is due to the assumption that the attacker has only access to the ampere-based \ac{ts}. If the attacker has access to voltage values, the corresponding features can not be extracted from the tail, as the tail corresponds to the constant voltage phase. Therefore features extracted from voltage values during constant voltage may be under-representative of the battery's behavior. If the attacker has access to both current and voltage \ac{ts}, current features can be extracted from tails, whereas voltage features can be extracted during the constant current phase.

By noticing that each battery follows the current limits imposed by the pilot differently, we generate a further \ac{ts} to be exploited to extract more features.  Together with the tail, in \emph{EVScout2.0} we exploit the Delta \ac{ts}, i.e., the \ac{ts} given by the punctual difference between the current \ac{ts} and the pilot \ac{ts} during the constant current phase. Delta \ac{ts} hence includes all the data from the beginning of the \ac{ts} up to the beginning of the tail. We believe that this derived \ac{ts} uniquely characterizes the behavior of each specific \ac{ev} as we will explain in Section~\ref{sec:delta}.

\subsection{Tail Identification}\label{sec:tailsIdent}

Charging sessions are not necessarily comprehensive of the constant voltage phase, as a user may need to leave before the full charge is reached. In~\cite{sun2020classification} the authors presented a framework to cluster similar charging behavior basing on the charging tail. We exploit this portion of the \ac{ts} to perform more detailed profiling. Since \emph{EVScout2.0} exploits the tails during the constant voltage phase, we exploit the algorithm we proposed in~\cite{brighente2021} to identify whether the considered session includes a constant voltage phase. The presence of a tail implies that the session terminates with full \ac{soc}, and eventually zero-current exchanged between \ac{ev} and \ac{evse}. Tails, however, can not be uniquely identified by the presence of zeros in the current \ac{ts}, as this may be due to idle phases during the power scheduling process at the grid side. Furthermore, scheduling may cause shot noise in the \ac{ts} also after full \ac{soc}, leading to spikes in the \ac{ts}. Therefore, we designed a suitable tail reconnaissance algorithm. 

\begin{figure}
    \subfigure[left][Raw current profile.]
    {\includegraphics[width = 0.45\columnwidth]{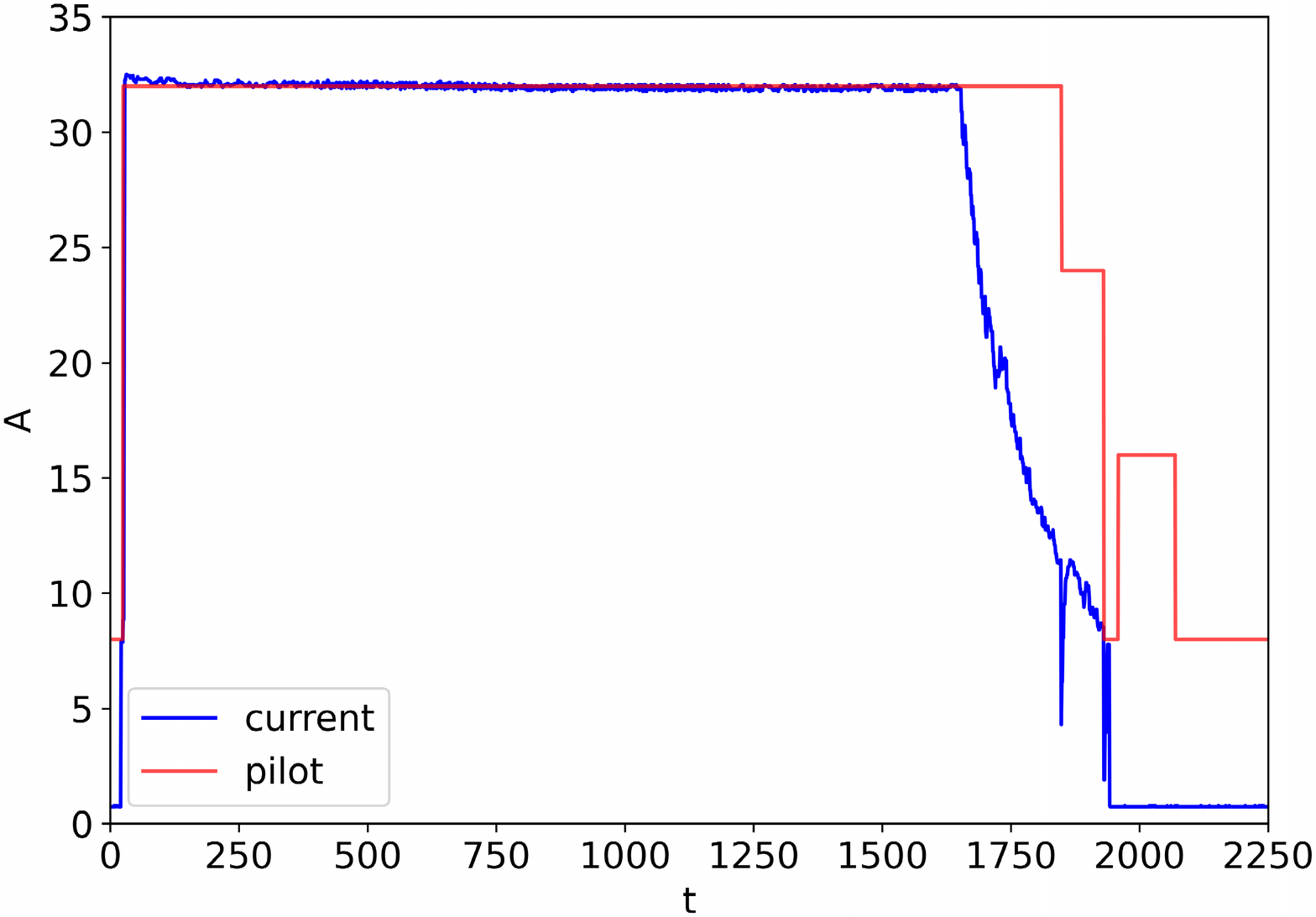}
    \label{fig:pc_normal}}
    \subfigure[right][Current filtered with a moving average filter with  $N_{avg}=25$.]{
    \includegraphics[width = 0.45\columnwidth]{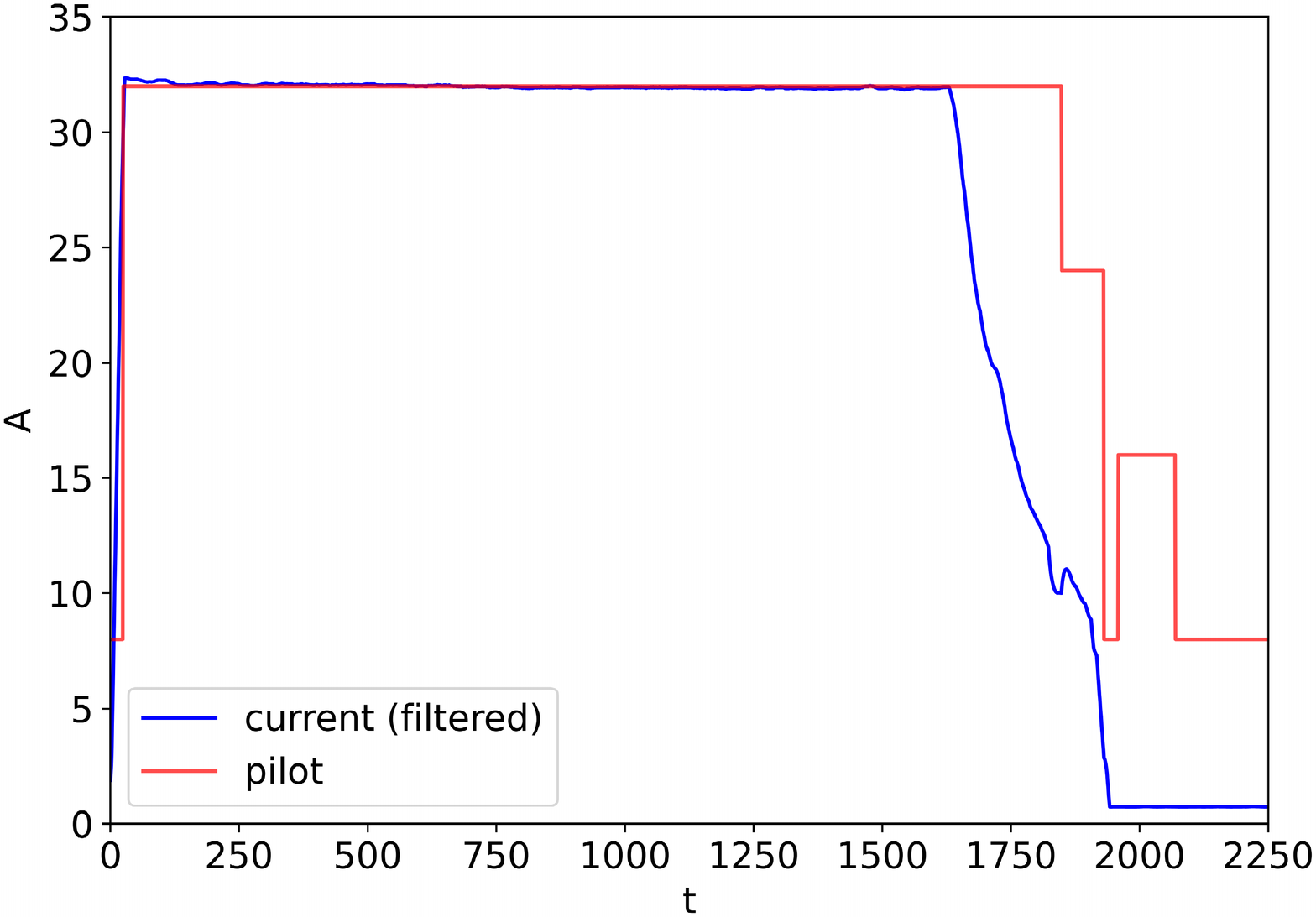}
    \label{fig:pc_filtered}}
 	\caption{The current (normal and filtered) with respect to the pilot during a sample charge.}
 	\label{fig:pilot_current_graph}
\end{figure}

To mitigate the aforementioned effects of scheduling and to highlight the trends in the considered \ac{ts}, we propose to apply a suitable filter. In particular, we filter both the current \ac{ts} and the pilot \ac{ts} with a length $N_{\rm avg}$ moving average filter. Given time instant $t$ and denoting the electric current value at time $t$ as $c(t)$, the output value $y(t)$ of the moving average filter at time $t$ is given by
\begin{equation}\label{eq:mAvg}
    y(t) = \frac{1}{N_{\rm avg}} \sum_{m=0}^{N_{\rm avg}-1}c(t-m).
\end{equation}
The effects of the moving average filter are shown in Figure~\ref{fig:pilot_current_graph}. We see that, with respect to the non-filtered current \ac{ts} in Figure~\ref{fig:pc_normal}, the current \ac{ts} in Figure~\ref{fig:pc_filtered} has a smoother behavior as the filter removes most of the noise and scheduling artifacts. Notice that different filter implementations can be considered, e.g., low pass filter. However, a low pass filter requires a more accurate design, and it also leads to ringing effects, which may be misleading for trend, and hence tail identification. 

If the filter has a sufficient length, its effects also include spikes removal. This eases the identification of tails in the \ac{ts}, as we can rely on the presence of steady zero values when the full charge is reached. In detail, if the current \ac{ts} assumes zero values from $t_{\rm start}$ up to its end, then we can assume that full \ac{soc} has been reached.
Tails are characterized by a descending trend in the \ac{ts}, as shown from the current behavior during the constant voltage phase in Figure~\ref{fig:soc} and verified in Figure~\ref{fig:pilot_current_graph}. By forward analysis of the current \ac{ts}, it is difficult to identify the time instant corresponding to the beginning of the tail, as this would imply the analysis of the overall \ac{ts}. Therefore, we propose to proceed backward from the point where full \ac{soc} is obtained. Proceeding from $t_{\rm start}$ backward, we identify the tail by accounting for the number of samples in the current \ac{ts} reporting an ascending trend. Notice that, even though scheduling and noise could affect the trend of the \ac{ts}, its effects are mitigated by the moving average filter (see Figure~\ref{fig:pc_filtered}). A perfectly backward-ascending trend is given by a negative difference between the values at time $t$ and $t-1$, i.e., $y(t)-y(t-1) < 0$. However, we notice that tails do not always exhibit a perfectly backward-ascending trend. In fact, if the non-filtered \ac{ts} is affected by heavy noise, its effects are still visible after filtering. Therefore, we relax the concept of perfect backward-ascending trend including samples for which $y(t)-y(t-1) \leq \varepsilon$, with $\varepsilon$ being a small positive value. Furthermore, we also allow for short descending trends by accounting for $T_{\rm max}$ consecutive segments for which $y(t)-y(t-1) > \varepsilon$. If this is the case for $T_{\rm max}$ consecutive samples, the trend is considered fully descending and hence discarded.

The steps of the tail extraction algorithm are shown in Algorithm \ref{alg:tail}. We denote as $\mathcal{I}$ the set of \ac{ev} indexes. Notice that each current \ac{ts} is associated with a unique pilot \ac{ts}. We denote as $\mathcal{C}(i)$ and $\mathcal{P}(i)$ the sets of the current and pilot \ac{ts} associated with ID $i$, respectively. We assume that the two sets are ordered such that the current and pilot \ac{ts} associated to a certain charging session are associated with the same index in the two sets. Therefore, we define set $\mathcal{W}(i) = \{\mathcal{C}(i),\mathcal{P}(i)\}$, whose elements $(c,p)$ are the couples of current and pilot time series respectively taken from sets $\mathcal{C}(i)$ and $\mathcal{P}(i)$. For each \ac{ts} $c \in \mathcal{C}(i)$, we compute the filtered current and pilot \ac{ts}, respectively denoted as $\tilde{c}$ and $\tilde{p}$, and we search for the steady zero values instant $t_{\rm start}$. If $t_{\rm start}$ is found, we proceed by identifying the number of samples of the tail. Given our definition of backward ascending trend, we compute the number of samples for which $\tilde{c}(t)-\tilde{c}(t-1) \leq \varepsilon$. As short descending trends are also allowed, we account for the number $n$ of consecutive descending samples, and if it exceeds $T_{\rm max}$ we stop the counter. However, if an ascending segment is found after a descending samples series, the counter is reset. Given the number $S$ of tail's samples, the tail $\tilde{c}(t_{\rm start},s)$ is obtained from the filtered current \ac{ts}, starting from $t_{\rm start}-s$ up to $t_{\rm start}$. The tail $\tilde{p}(t_{\rm start},s)$ associated to the pilot \ac{ts} is analogously obtained, starting from $t_{\rm start}-s$ up to $t_{\rm start}$. Both current and pilot \ac{ts} are eventually added respectively to the set $\mathcal{T}_c(i)$ and $\mathcal{T}_p(i)$ of current and pilot \ac{ts} tails associated with \ac{ev} ID $i$.

\begin{algorithm}[t]
	\SetAlgoLined
	\KwData{$\mathcal{W}$, $\mathcal{I}$, $C_{\rm max}$, $T_{\rm max}$, $N_{\rm avg}$ }
	\KwResult{$\mathcal{T}_c$,$\mathcal{T}_p$ }
	\For{$i \in \mathcal{I}$}{
		\For{$(c,p) \in \mathcal{W}(i)$}{
		    compute $\tilde{c}$ and $\tilde{p}$ via (\ref{eq:mAvg})\;
		    compute $t_{\rm start}$\;
		    \eIf{$t_{\rm start}$ found}
		    {$n = 0$, $s=0$ \;
		    \For{$t=t_{\rm start}, t_{\rm start}-1, \ldots, 1 $}{
	            \eIf {$\tilde{c}(t)-\tilde{c}(t-1) > \varepsilon$}{
	    	        $n=n+1$\;
	                }{
	                $n=0$\;
	                $s = s+1$\;
	                }
	            \If{$n = T_{\rm max}$}{
	                exit loop\;
	                }     
	            }
	         $\mathcal{T}_c(i) = \mathcal{T}_c(i) \cup \tilde{c}(t_{\rm start},s)$\; 
	         $\mathcal{T}_p(i) = \mathcal{T}_p(i) \cup \tilde{p}(t_{\rm start},s)$\; 
	        }
	        {go to next $c$\;}
	        }
	        
	    }
	\caption{Tail extraction algorithm.}
	\label{alg:tail}
\end{algorithm}

\subsection{Delta TS computation}\label{sec:delta}

Aside from current tails, we compute another \ac{ts} to be used for extracting features for the classifiers. Since different batteries' charging sessions have different charging parameters, the maximum current which can be absorbed is variable. Furthermore, pushed by advanced charging algorithms~\cite{Lee2019acndata}, during some periods, the \ac{ev} can be forced into charging at a lower current, e.g., to deal with peak leveling during busy hours.
However, as visible in Figure~\ref{fig:pilot_current_graph}, generally, the battery does not charge at the exact amount of energy expected from the pilot signal. Furthermore, the absorbed current often exhibits some small variations around the maximum current deliverable by the charging column. This is particularly true when considering the behavior of the two \ac{ts} is the time preceding the tail. In order to capture these changes, we compute \textit{Delta \ac{ts}}, i.e., the \ac{ts} given by the combination of the current \ac{ts} and the control pilot \ac{ts} during the constant current phase (i.e., the period preceding the tail).

To compute Delta \ac{ts}, we calculate the difference between the current and pilot \acp{ts} at each time instant during the constant current phase. While the pilot \ac{ts} generally is not affected by noise, the current \ac{ts} instead exhibits some tiny positive and negative spikes, as shown in Figure~\ref{fig:pc_normal}. While the tails generally follow a decreasing trend, the values assumed by the \acp{ts} before the tail (i.e., the ones used to compute the Delta \ac{ts}) are generally more constant. Since the moving median filter provides better performance in removing noise and spikes when the data in the neighborhood of the peak are quite constant~\cite{NRajeshKumar2015}, we use it instead of the moving average filter used in Section~\ref{sec:tailsIdent}.    

Given time instant $t$, we denote the electric current value at $t$ as $c(t)$ and the correspondent pilot value as $p(t)$. The resulting point at time $t$ in the Delta \ac{ts} can be expressed as:
\begin{equation}\label{eq:delta_current}
    z(t) = p(t) - median \left (c \left [t-\left \lfloor{\frac{
    N_{\rm avg}}{2}}\right\rfloor, t + \left \lceil{\frac{N_{\rm avg}}{2}}\right\rceil \right ] \right ).
\end{equation}
where $c[a, b]$ represents the array of values of the \ac{ts} $c$ from $t = a$ to $t = b$, $N_{\rm avg}$ is the filter length, and $median(x)$ is the median value of array $x$ (i.e., the middle value separating the grater and lower halves of $x$).

\subsection{Improved Feature Extraction}\label{sec:featExt}

Segmentation represents a classical approach for feature extraction in \acp{ts}~\cite{deng2013time,christ2016distributed,conti2016mind}. Unfortunately, since the \acp{ts} we consider here are generally short with no stationary components, segmentation is not a viable solution. Therefore, we do not further process tails before extracting features. 
In our previous work~\cite{brighente2021}, we computed mean, mode, median, max value, standard deviation, auto-correlation, length of the tail, and the slope of the linear approximation for each tail. Furthermore, we used as a feature the total kW delivered and the overall session time duration, leading to a total of $18$ features considering both pilot and current \acp{ts} independently. Instead, in this work, we adopted a more sophisticated feature extraction process that can extract several more features. 

We analyze different widely used feature extraction tools which can automatically extract features from a given \ac{ts}~\cite{barandas2020tsfel, morchen2003time, deng2013time}. We use Time Series Feature Extraction based on Scalable Hypothesis tests (\texttt{tsfresh})~\cite{tsfresh}, which is available as a Python package easily integrable with other tools such as Scikit-learn~\cite{scikitlearn}. \texttt{tsfresh} exploits the power of $63$ \acp{ts} characterization methods to extract hundreds of features from a \ac{ts}. Moreover, it contains functions to slightly reduce the number of features to remove the less meaningful ones.  
We use \texttt{tsfresh} to extract features from the current tails and the delta current-pilot \acp{ts}. It is worth mentioning that, with respect to the previous work~\cite{brighente2021}, we decided not to employ features extracted from the tails of the control pilot \ac{ts} since the pilot tail is generally based on the optimization algorithm~\cite{Lee2019acndata} and not on the behavior of the battery. For example, even if the battery has reached its maximum \ac{soc}, the control pilot could remain at a high value if the grid has energy available. In the same way, if many vehicles start requesting energy, the control pilot will reduce its value independently of the \ac{soc} of our target \ac{ev}. Furthermore, we removed the needs on the duration of the charge and the total energy absorbed by the battery (kW) since they can depend on the user behavior and the \ac{soc} of the battery at the beginning of the charging process.

Since \texttt{tsfresh} can generate around $800$ features for each \ac{ts}, we removed those that are not relevant for our classification problem. To reduce the number of features, hereinafter denoted as $NoF$, we select the most significant ones by using \texttt{SelectKBest} of Scikit-learn~\cite{scikitlearn} with the \texttt{chi2} function, which is suitable for classification purposes. 
Since the chi-squared measures the dependence between stochastic variables, we can highlight and then select only the features that offer more information for the classification. We employ this strategy with respect to other more complex methods such as Random Forest feature reduction since \texttt{SelectKBest} can achieve approximately the same results with lower computational complexity.

\section{Evaluation Framework Description}\label{sec:evaluation}

In the following, we present the experiments we use to test \emph{EVScout2.0}. In particular, in Section~\ref{subsec:dataset} and Section~\ref{subsec:new_dataset} we describe respectively the ACN Infrastructure and Dataset on which we based our analysis. We then explain how the new version of the dataset differs from the one in the previous work. Then, in Section~\ref{subsec:alg_test} we outline the machine learning classifiers we use to profile \acp{ev}.

\subsection{The ACN Infrastructure and Dataset}\label{subsec:dataset}

In order to test \emph{EVScout2.0}, we exploit the \ac{acn} proposed in~\cite{Lee2019acndata}.
It consists of level $2$ \acp{evse} connected with a central controller that regulates power exchanges in the grid. Employing an online optimization framework, the \ac{acn} allows adapting the power exchanged in the grid, satisfying users' power demand while coping with the grid's capacity limits. The dataset comprises charging sessions from actually deployed \acp{acn}, each reporting user-specific measurements such as the arrival and departure time, the kw/h delivered, current and pilot \acp{ts} collected between the \ac{ev} connection and disconnection time. Notice that, although the user may have planned for a full recharge during the selected period, this may not be reflected in the \ac{ts}. In fact, due to the variable number of connected \acp{ev}, the upper power limit of the grid, and the premature departure of the user, the battery may not be fully charged at disconnection time. Notice also that, in the \ac{acn} dataset, not all \ac{ts} are sampled with the same period. However, we avoid upsampling with the filtering because it can introduce statistical features that are not representative of the analyzed battery. Each user in the dataset is identified by a unique ID, which is associated with the owned \ac{ev}.

\subsection{New Dataset}\label{subsec:new_dataset}

Like in our previous work~\cite{brighente2021}, we decided to use the \ac{acn} dataset. However, since the dataset was recently enlarged, we expanded the number of \acp{ev} considered from $22$ to $187$. To generate the dataset, we selected all the available \acp{ev} up to June 18, 2021, from the \texttt{caltech} site (one of the three locations available in the dataset). We also excluded by default \acp{ev} without charges and charges without any \ac{ev} assigned (i.e., anonymous charges). To download the dataset, we employ the Python APIs provided by the ACN Dataset~\cite{Lee2019acndata}. The number of charges associated with each \ac{ev} ranges from one to over $300$. However, not all the charges contain a tail to be analyzed. We therefore remove the corresponding \acp{ts} and hence \acp{ev}. We consider all the \acp{ev} with more than $8$ employable charging processes associated. We made this choice to be able to effectively use cross-validation even for those \acp{ev} with a small number of charges. After this cleanup, $137$ \acp{ev} remains in the dataset, resulting in a dataset more than six times bigger than the one used to test EVScout in~\cite{brighente2021}.

\subsection{Classification Algorithms Comparison}\label{subsec:alg_test}

The first step of \emph{EVScout2.0} is to build a suitable dataset to be exploited for profiling, as explained in Section~\ref{subsec:new_dataset}. It is worth mentioning that the dataset does not provide any information regarding the make and model of the analyzed \acp{ev}. Since the energy behavior highly depends on the chemical reactions of the single battery, we think that \emph{EVScout2.0} could be able to distinguish \acp{ev} of the same model. Furthermore, the batteries employed by the analyzed \acp{ev} all belong to the same class, i.e., constant current/constant voltage. However, since both classes discussed in Section \ref{sec:sys_threat} show particular behaviors in time, we believe that our attack could be easily extended to the constant power/constant voltage class. The effectiveness of \emph{EVScout2.0} in these cases will be investigated in future works.

For each session, \emph{EVScout2.0} first identifies whether a tail is present and discards all the other sessions. Then it builds a feature vector for each tail, associating it with the ID of its corresponding \ac{ev}. We test \emph{EVScout2.0} across all \acp{ev} in the dataset by averaging the performance obtained with every single classifier. In particular, we implement a binary classifier for each \ac{ev}, and we associate each feature vector of the target \ac{ev} with label $1$, otherwise with label $0$. The overall performance of the obtained classifiers is averaged considering $100$ randomly created training and testing sets, except \ac{rf} and \ac{ada} classifiers for which, for timing reasons, we consider $25$ iterations. The overall performances of \emph{EVScout2.0} are obtained by averaging the results obtained for each ID's classifier. 

Let us denote as $Q$ the ratio between the number of feature vectors associated with the target \ac{ev} and the number of feature vectors associated with other \acp{ev}. Hence, $Q$ measures the amount of unbalancing in the considered dataset. In order to further assess the performance of \emph{EVScout2.0}, each classifier is tested for multiple $Q$ values. Regardless of the value $Q$, the $80\%$ of the dataset has been used for training and the remaining $20\%$ for testing, unless otherwise specified.
As the number of feature vectors of a single \ac{ev} is smaller than the overall number of feature vectors, when considering small $Q$ values, the set of feature vectors associated with other \acp{ev} is randomly created from the overall set. Another value that can lead to different results is the number of features $NoF$ employed in the classification. Since our feature extraction strategy returns about $1500$ features, using them all can lead to overfitting. We show in the next sections how different $NoF$ values affect the classification performance and provide a justification for the choice of a suitable $NoF$ value that provides a good threshold to balance classification performance and overfitting.



\section{\emph{EVScout2.0} Performance: Vehicle Profiling}\label{sec:profiling}

We first assess the performance of \emph{EVScout2.0} in terms of profiling every vehicle based on its charging behavior. To this aim, we leveraged and compared different common machine learning algorithms for classification. We describe in Section~\ref{subsec:classification} the exploited classifiers and provide a discussion on their hyper-parameters setting. Then, in Section~\ref{subsec:prof_results}, we present the results obtained via \emph{EVScout2.0} with the different classifiers.

\subsection{Classification Algorithms}\label{subsec:classification}

Once features have been identified and selected, \emph{EVScout2.0} feeds them to a binary machine learning classifier.
The profiling task can be formulated as a supervised classification problem, where a two-class classifier is trained with both features from the target \ac{ev} and features from all other \acp{ev}. In particular, we assume that a classifier whose input is the features vector from the target \ac{ev} shall return output value $1$, otherwise it shall return output value $0$. 

We evaluate our pipeline by using and comparing six different common machine learning models, namely:
\begin{itemize}
    \item Support-Vector Machine (SVM) classifier~\cite{suykens1999least};
    \item k-Nearest-Neighbours (kNN) classifier~\cite{guo2003knn};
    \item Decision Tree (DT) classifier~\cite{breiman1984classification};
    \item Logistic Regression (LR) classifier~\cite{hosmer2013applied}
    \item Random Forest (RF) classifier~\cite{breiman2001random};
    \item ADA Boost (ADA) classifier~\cite{freund1997decision}.
\end{itemize}

Hyper-parameters optimization is obtained via grid search with cross-validation. Table~\ref{tab:res} indicates the different parameters employed in the grid search for each model. The training set is suitably divided into training and validation sets, which we test on a grid of possible hyper-parameters. 
Notice that all six classifiers are standard machine learning algorithms without deep architectures. Although deep learning automates the feature extraction process, a large number of samples shall be used to train deep architectures effectively. The use of non-deep structures allows us to show the feasibility of \emph{EVScout2.0} over our currently available dataset. The same motivation resides behind the choice of binary classifiers. In fact, a single multi-class classifier can be designed to have a single class for each \ac{ev}. However, multi-class classifiers require a larger dataset for training purposes compared to binary classifiers. Although we consider a larger dataset with respect to our previous work~\cite{brighente2021}, it is not big enough to provide interesting results with deep models or a multi-class scenario.

\begin{table}[t]
\centering
\begin{tabular}{lll}
\hline
\textbf{Model} & \textbf{Parameter} & \textbf{Values} \\ \hline \\[-2ex]
\multirow{3}{*}{SVM} & kernel            & {[}`rbf'{]} \\ & regularization (c)            & $[1, 10, 10^{2}, 10^{3}]$ \\
    & gamma ($\gamma$)     & $[10^{-4}, 10^{-3}]$ \\[-2ex] \\ \hline \\[-2ex]
\multirow{3}{*}{kNN} & n\_neighbors & $[1, \ldots , 10]$ \\
& weights            & {[}`uniform', `distance'{]} \\
& metric     & {[}`euclidean', `manhattan'{]}  \\[-2ex] \\ \hline \\[-2ex]
\multirow{2}{*}{DT} & criterion & {[}`gini', `entropy'{]} \\
& max\_depth            & $[8, 10, 14, 30, 70, 110]$ \\[-2ex] \\ \hline \\[-2ex]
\multirow{2}{*}{LR} & max\_iter & $[5000]$ \\
& regularization (c)            & $[10^{-2}, 1, 10^{2}]$ \\[-2ex] \\ \hline \\[-2ex]
\multirow{2}{*}{RF}  & n\_estimators & $[50, 200, 1000]$ \\
    & max\_depth   & $[10, 100, None]$   \\[-2ex] \\ \hline \\[-2ex]
ADA  & n\_estimators & $[10,100,500,1000,5000]$ \\[-2ex] \\ \hline \\[-2ex]
\end{tabular}
\label{tab:res}
\caption{Grid search cross validation parameters for each model.}
\end{table}

\subsection{Profiling Results}\label{subsec:prof_results}

We exploit the implementation of the classification algorithms in~\cite{scikitlearn}. Results are assessed in terms of precision $P$, recall $R$, and $F1$. We present numerical results assessing the validity of \emph{EVScout2.0} as a function of $Q$, the amount of unbalancing in the dataset. Since traditional performance measures such as $F1$ may be misleading when considering highly unbalanced datasets, the geometric mean (G-Mean) between recall and specificity has been proposed as a suitable performance metric~\cite{ferri2009experimental}. Therefore, we consider G-Mean as a proper indicator of the validity of \emph{EVScout2.0} for large $Q$ values. By denoting as $TP$, $TN$, $FP$ and $FN$ respectively the number of true positive, true negative, false-positive and false-negative outcomes, we can express the recall as $R = \frac{TP}{TP + FN}$, and specificity as $\alpha =\frac{TN}{FP + TN}$. G-Mean is hence obtained as $\text{G-Mean} = \sqrt{\alpha R}$.

Since the tail extraction process is automated, the number of extracted tails also depends on the parameter values. Basing on our previous work~\cite{brighente2021}, we set $N_{avg}$ (i.e., the size of the moving average filter) to $25$, which was the best value in terms of classification scores. We used the same value for the median filter to maintain coherence in the \acp{ts} since, albeit the necessary differences, the two filters are quite similar.

We employed the implemented versions of the classifiers provided by the Scikit-Learn Python library~\cite{scikitlearn}. For each model, we employed a \texttt{GridSearchCV} to tune the model using grid search with cross-validation. In the following, we present an analysis of the results with respect to different parameters and values.  

\subsubsection{\textbf{Number of Features \emph{NoF}}}

\begin{figure}[t]
    \centering
    \includegraphics[width=0.6\columnwidth]{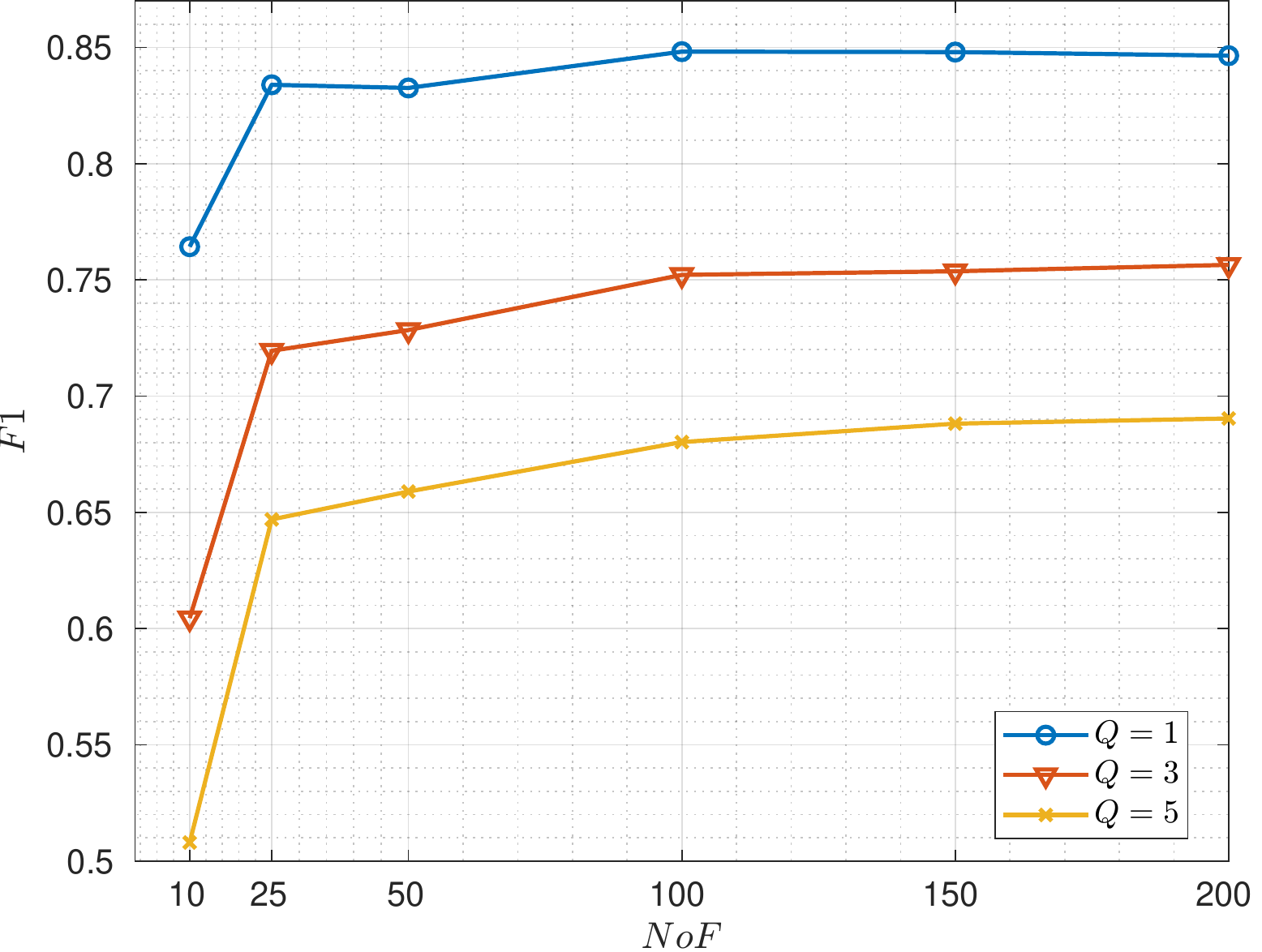}
    \caption{$F1$ score of \ac{knn} classifier with different numbers of features $NoF$s.}
    \label{fig:features_numbers}
\end{figure}

Firstly, we provide an analysis of the scores based on the number of features $NoF$ maintained after the feature extraction phase. In Figure~\ref{fig:features_numbers} we plot the $F1$ scores for different ratios $Q$ using the same model (\ac{knn}) but varying the numbers of features $NoF$ from $10$ to $200$. We can see a significant increase in the score going from $10$ to $25$ features, especially for higher ratios $Q$. From $100$ features up, the increase is instead negligible. For this reason, we selected $NoF=100$ for the other experiments in this paper, unless otherwise specified.     

\subsubsection{\textbf{Unbalance of the Dataset \emph{Q}}}

To understand how our models deal with the unbalance of the dataset (i.e., the ratio between the number of feature vectors associated with the target \ac{ev} and the number of feature vectors associated with other \acp{ev}), we test all the six models against different values of $Q$. We test $Q$ values ranging from $1$ (i.e., the same number of feature vectors for target and for the other \acp{ev}) to $5$ (i.e., five times more features vectors not related to the target \ac{ev}). 

\begin{figure}[t]
    \subfigure[b][precision]
    {\includegraphics[width = 0.45\columnwidth]{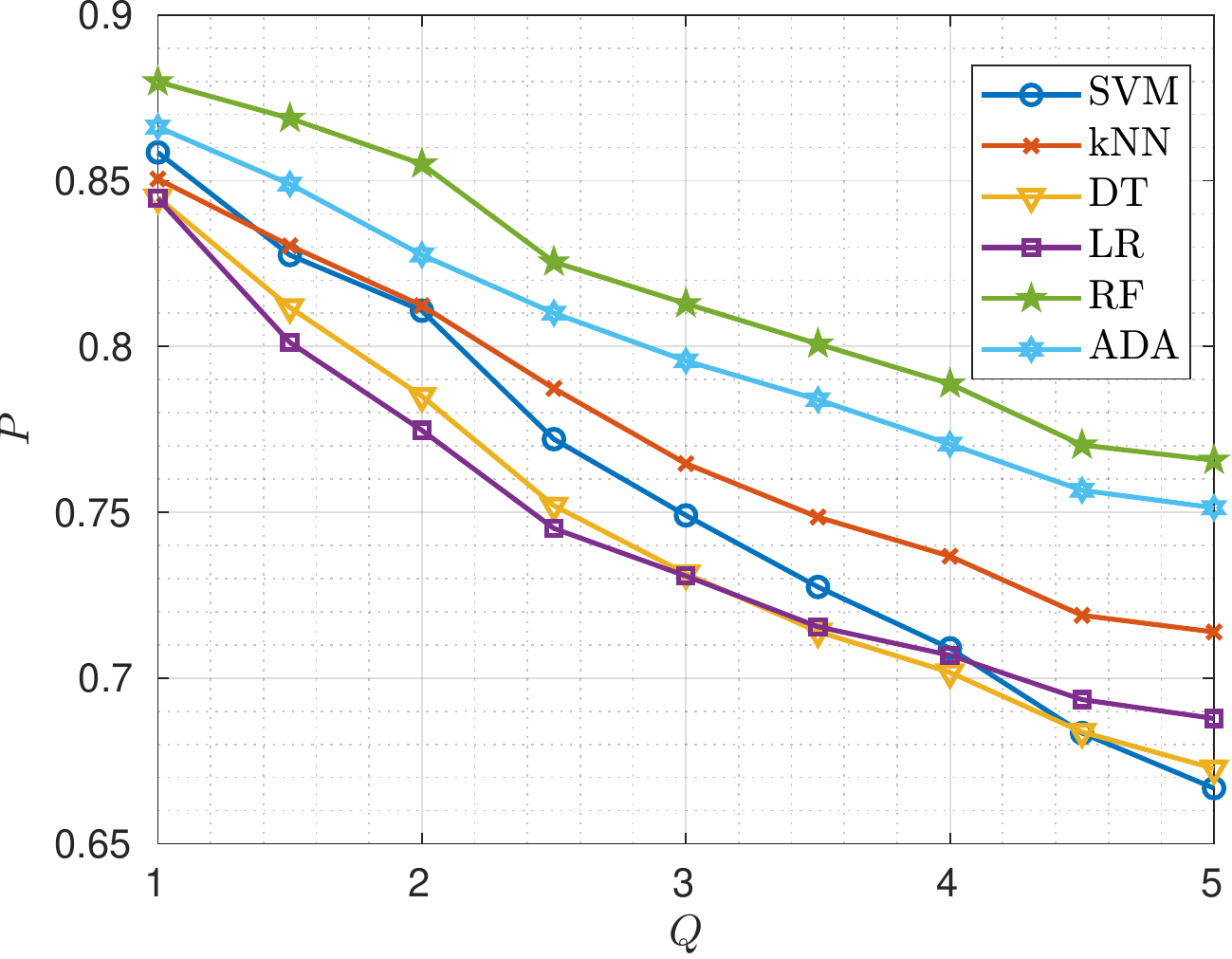}
    \label{fig:pQ}
    }
    \subfigure[b][recall]{
    \includegraphics[width = 0.45\columnwidth]{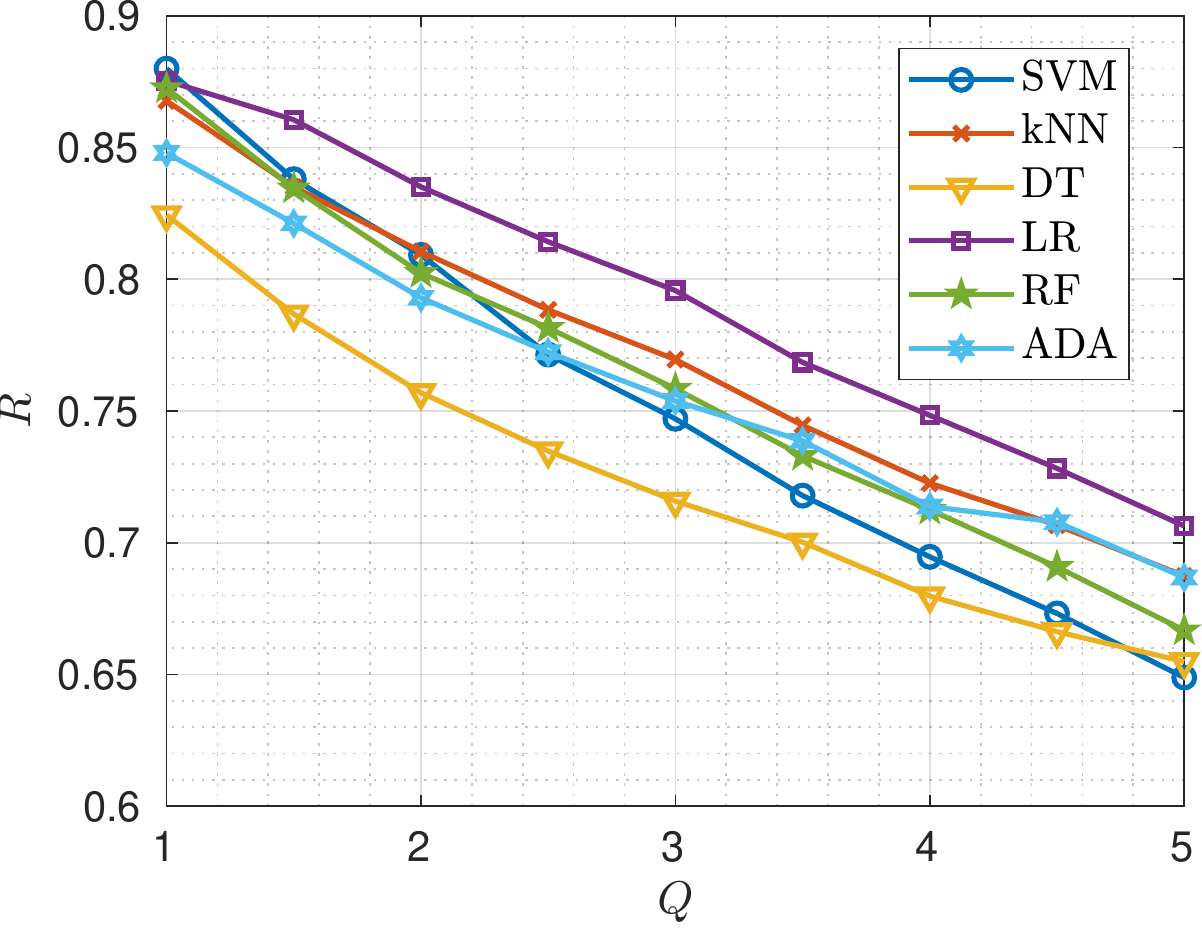}
    \label{fig:rQ}
    }
    \subfigure[b][$F1$]
    {\includegraphics[width = 0.45\columnwidth]{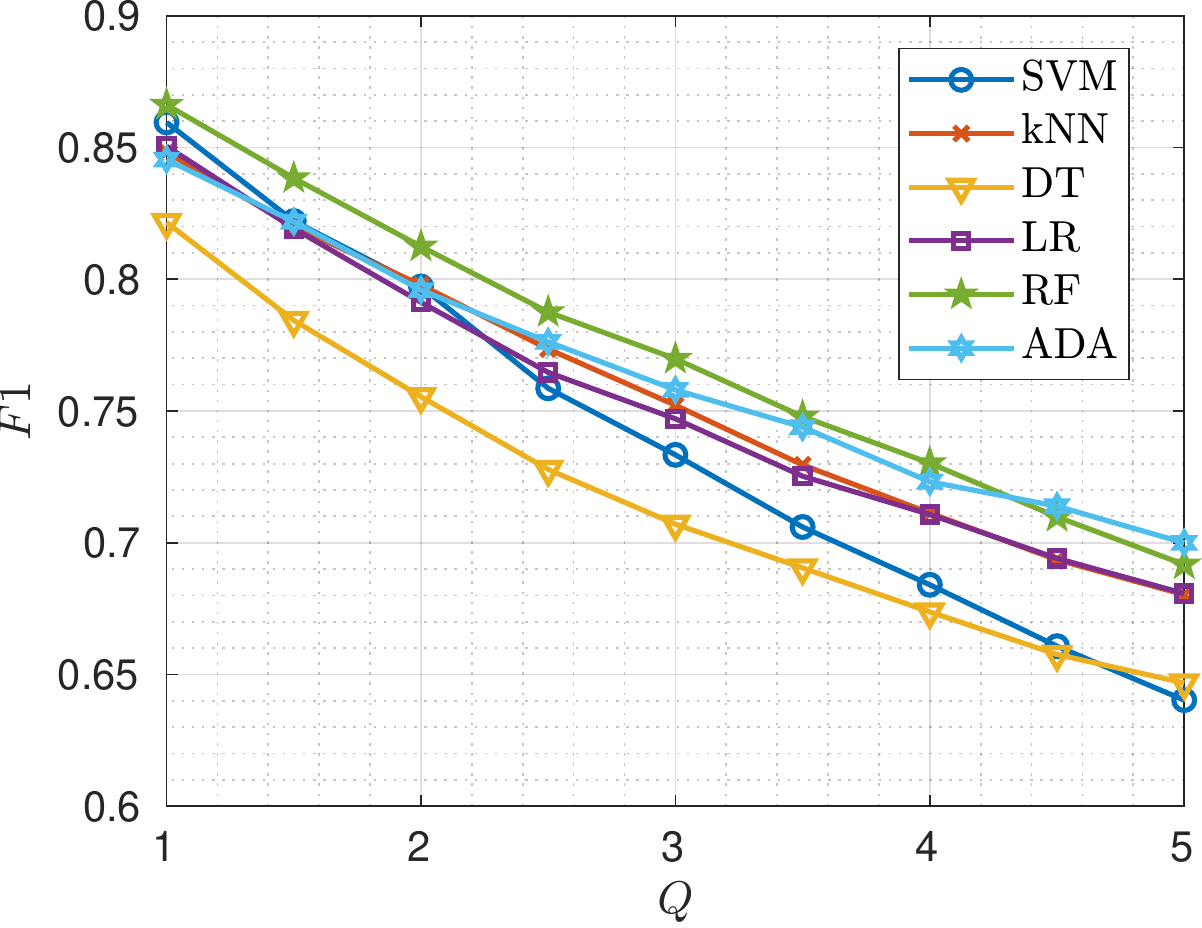}
    \label{fig:f}
    }
    \subfigure[b][G-Mean]{
    \includegraphics[width = 0.45\columnwidth]{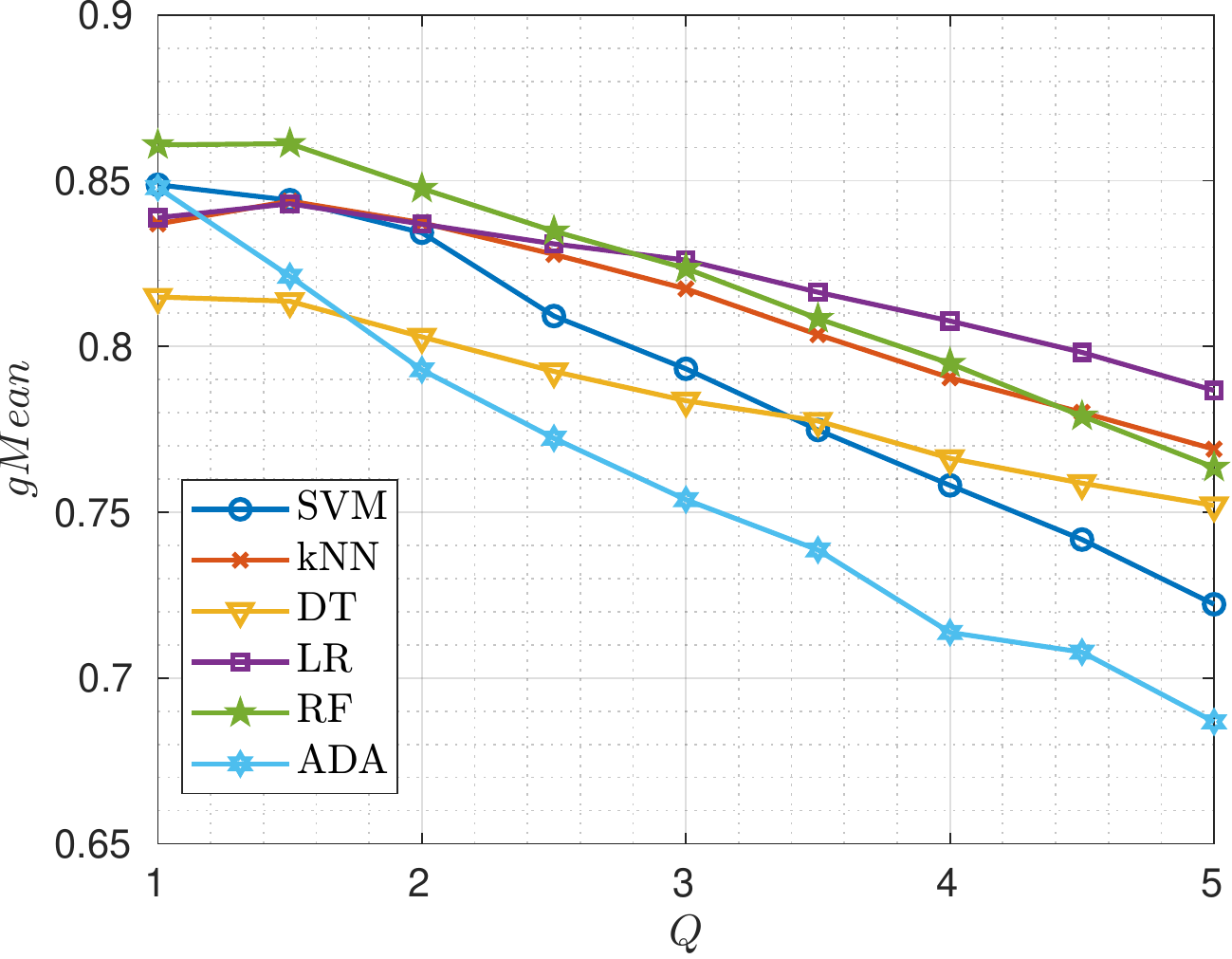}
    \label{fig:g}
    }
 	\caption{Performance of \emph{EVScout2.0} for different amount of unbalancing in the dataset. Results are shown for the different classifier algorithms and $NoF=100$. We see that good classification performance are obtained for all classifier. As $Q$ increases, some classifiers are more robust than others to increasing unbalancing. Results on G-Mean show that profiling can also be achieved in largely populated networks.}
 	\label{fig:q_values}
\end{figure}

Figure~\ref{fig:q_values} shows the results obtained by \emph{EVScout2.0} for different $Q$ values. It is possible to notice how all the scores decrease with $Q$ for all six models. As the number of considered \acp{ev} increases with $Q$, the chance of two users having the same \ac{ev} model or having \acp{ev} with similar charging profiles increases. This is reflected in a worsening of classifiers' performance. With no unbalancing (i.e., $Q=1$), we reach the highest precision of $0.88$ with the \ac{rf} classifier. On the other hand, almost all the classifiers reached at least $0.86$ in recall, with the exception of \ac{dt} which has the lower performances almost for every indicator. We notice that the \ac{rf} classifier is the most resistant to changes in $Q$ if we look at the precision, while \ac{lr} offers the best recall scores. If instead we look at both the scores, even if the absolute values are generally lower, \ac{knn} and \ac{ada} seem to be the most resistant to higher $Q$. Furthermore, we can observe that for the maximum $Q=5$, precision and recall are respectively $0.77$ (with \ac{rf}) and $0.71$ (with \ac{lr}), meaning that \acp{ev} can still be profiled with sufficient confidence.

As for the other scores, also $F1$ degrades for increasing values of $Q$ for all the models. We can see that \ac{rf} is the best one for almost all the ratios $Q$, while for the highest two (i.e., $Q=4.5$ and $Q=5$) \ac{ada} performed slightly better. However, this is not the case in terms of G-Mean, confirming its validity for unbalanced datasets. In particular, \ac{ada} presents a large variance in terms of G-Mean, a sign that is not a suitable algorithm for highly unbalanced datasets. Other algorithms, such as \ac{rf}, \ac{knn}, and \ac{lr} are instead able to maintain high values of G-Mean for every value of $Q$. This shows that profiling can be achieved with good results irrespective of the amount of unbalancing in the dataset, i.e., a single user can still be profiled based on its charging profile also in largely populated networks.

However, all this consideration must be taken into account when designing \emph{EVScout2.0}. If the dataset distribution is known, the best algorithm for each case can be selected. In particular, \ac{rf} is advisable for perfectly balanced datasets, \ac{lr} for highly unbalanced datasets. Instead, if the dataset is unknown and a resilient model is needed, \ac{knn} can be a good choice. In fact, we employed \ac{knn} in many experiments hereinafter, also thanks to its fast training time with respect to other models such as \ac{rf} or \ac{ada}.

\section{\emph{EVScout2.0} Performance: Additional Performance Analysis}\label{sec:additional_analysis}

In addition to the classification-based analyses, we included additional scenarios based on the training set characteristics. Since the data conditions may be different in a real-world scenario, in the following, we propose an analysis taking into consideration different data properties.
In particular, in Section~\ref{subsec:training_variation} we examine different training set size values to assess the minimum number of charges needed to obtain sufficiently high classification scores. In fact, in a real-world attack, a large number of labeled traces may be challenging to obtain. 
In Section~\ref{subsec:battery_degradation}, we investigate if and how much the degradation of the Li-ion battery has an impact on the performance of \emph{EVScout2.0}. This analysis can be useful to understand how a model can still be precise when dealing with natural phenomena like physical battery degradation. 

\subsection{Training Size Variation}\label{subsec:training_variation}

In a real-world scenario, an attacker may be limited by the number of labeled charges she is able to get for a target vehicle. For instance, the attacker may not want to leave its malicious device in the field too much to reduce the possibility of being detected. 
To simulate this scenario, we analyze the performance of \emph{EVScout2.0} while varying the train set size. We avoid using as target \acp{ev} with less than $70$ charges with tails, while to create the group with the others \acp{ev} we employed the whole dataset. As a testing set, we always use the last $20\%$ of the available charges. For the training set, we set fixed values from the $80\%$ to the $10\%$ of the min number of charges for each \ac{ev} (i.e., $70$). In other words, the training set size ranges from $7$ to $56$ feature vectors for each \ac{ev}, with steps of $7$ charges.

\begin{figure}
    \subfigure[left][Training set size changing.]
    {\includegraphics[width = 0.45\columnwidth]{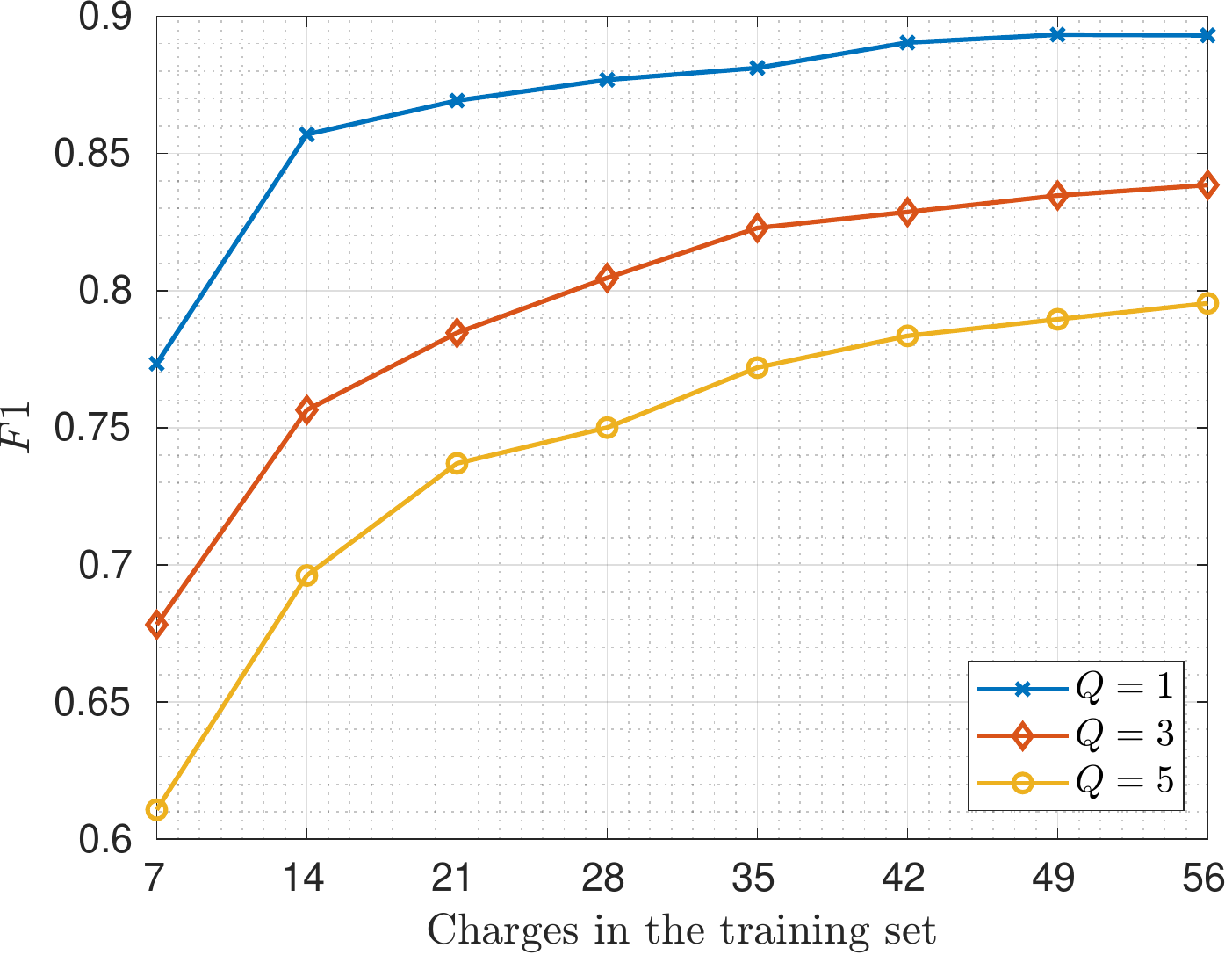}
    \label{fig:tr}}
    \subfigure[right][Changing $NoF$ and training set size for $Q=1$.]{
    \includegraphics[width = 0.45\columnwidth]{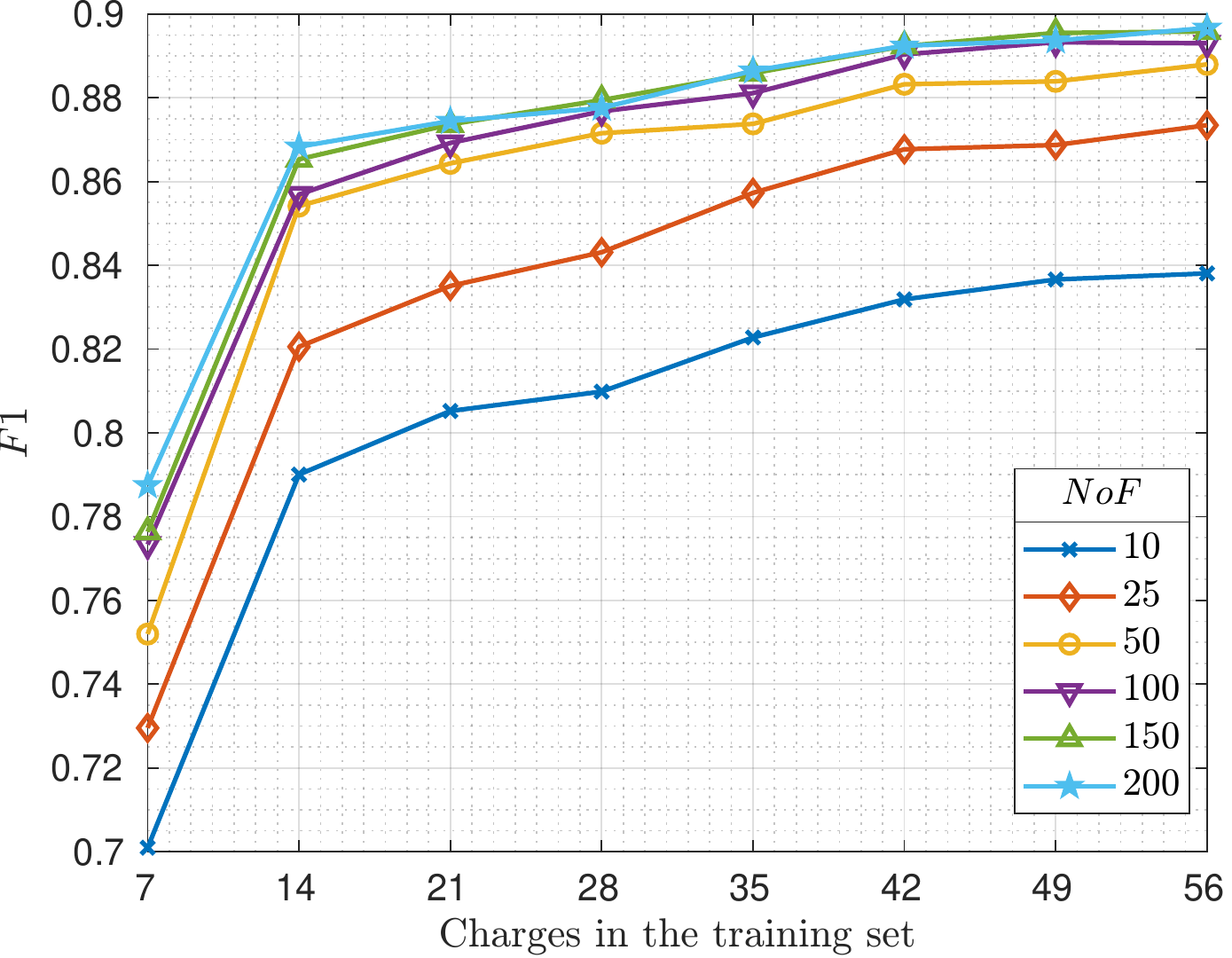}
    \label{fig:tr_features}}
 	\caption{$F1$ scores of \ac{knn} classifier while reducing the training set size.}
 	\label{fig:train_reduction}
\end{figure}

In Figure~\ref{fig:tr} we can see how the $F1$ decreases while decreasing the training set size. However, we can appreciate a significant fall only for the smallest values of the training set, while for training set sizes bigger than $14$ charges the increase is less relevant. Over $42$ charges, the performance increase is almost negligible, especially considering the smaller $Q$ ratios. We can also look at the absolute values of the $F1$. It is greater than $0.69$ also for the most unbalanced dataset when using $14$ feature vectors, showing discrete classification performance also with small training sets.

Furthermore, we also analyzed the impact of both training set reduction and features reduction. In Figure~\ref{fig:tr_features} we see the different behavior of the $F1$ score while varying the number of features and the training set size (we show the data only for $Q=1$ for clarity). As expected, there is a clear drop in the performances for the lower training set sizes. However, this can be partially compensated by employing a bigger number of features up to $100$. Over this threshold, the gain is negligible, coherently with what is presented in Section~\ref{subsec:prof_results} and Figure~\ref{fig:features_numbers}.  

\subsection{Battery Degradation Performance}\label{subsec:battery_degradation}

The degradation of a Li-ion battery used in an \ac{ev} is a topic widely discussed in literature~\cite{Pelletier2017}. During a battery life, many aspects can adversely affect its performance depending on the number of charge cycles, aging, operating, and storing conditions. Degradation on \ac{ev} batteries can lead to a shorter traveling distance and a reduced battery's available power output~\cite{Barre2013}. Generally, a battery is considered to be at the end of life when it already lost $20\%$ of its initial capacity. Modern Li-ion batteries should have a five to ten-year calendar life depending on the materials and how they are managed. They should be able to supply between $1000$ and $2000$ cycles of charge. However, many behaviors such as keeping the battery at a high \ac{soc} or high temperature, or often employing fast chargers, can reduce even more the lifetime of the battery~\cite{VETTER2005269, Barre2013}.

Being aware of this problem, we investigate if the battery degradation affects our profiling model. Since the charge profile of an \ac{ev} will always be different due to the variation of the physical properties of the battery, we analyzed how the extracted features degrade the performance of \emph{EVScout2.0} in time. To test how our model behaves in this context, we selected the $10$ \acp{ev} with the higher number of charges (i.e., more than $150$ charges for each \ac{ev}) spanning about two years. We train the \ac{knn} classifier in the first part of the data, and we then test it over multiple consecutive test set batches. In particular, the training set accounts for the \ac{ts} measured in a specific period and represents the baseline to measure the successive battery degradation. Each testing set is given by batches of $Z$ chronologically consecutive \acp{ts}, with $Z=5\%$ of the total available testing data. In other words, we considered as each test set a sliding time window of consecutive \acp{ts}. 

\begin{figure}
    \subfigure[left][Train set size $30\%$ of the data.]
    {\includegraphics[width = 0.45\columnwidth]{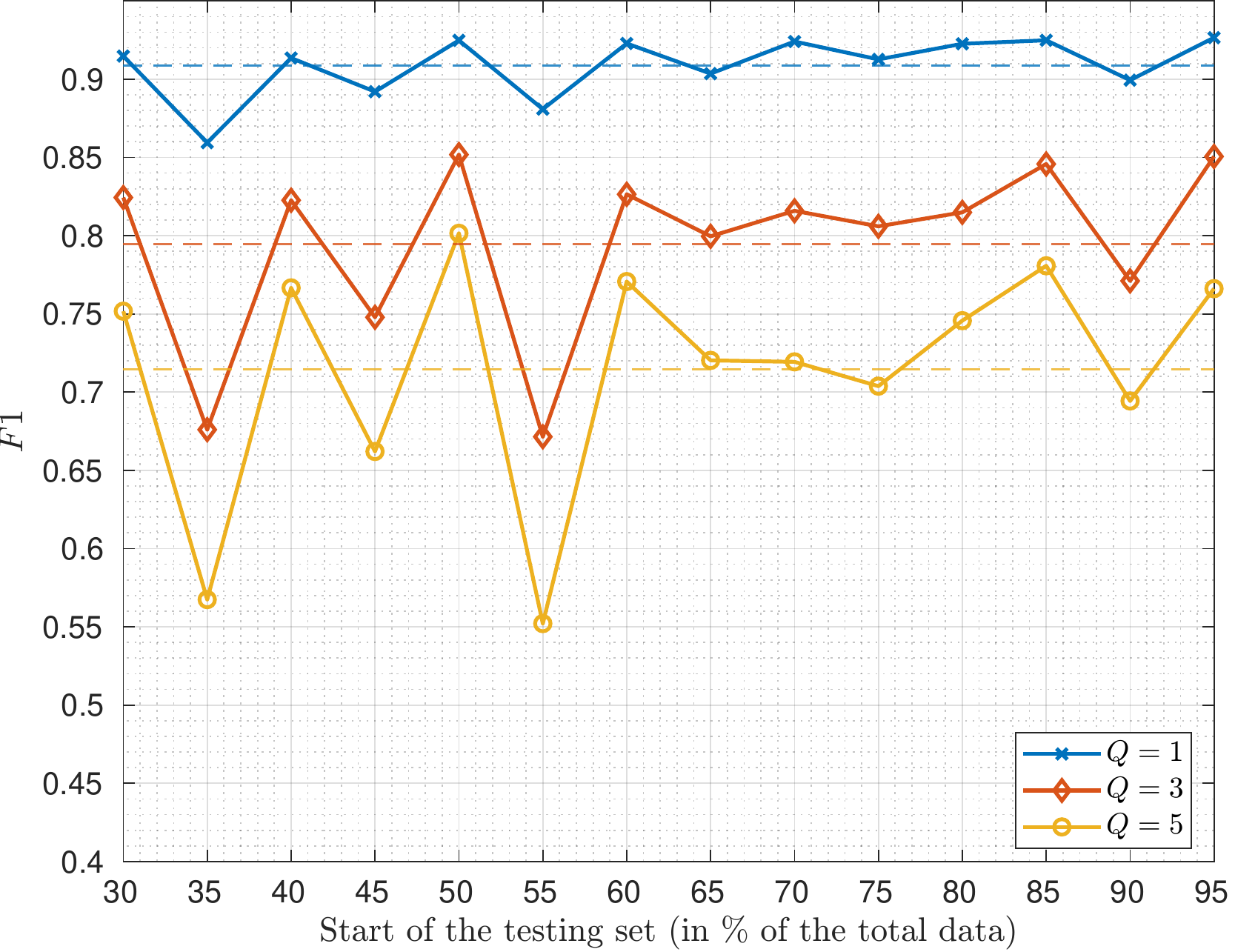}
    \label{fig:deg_3}}
    \subfigure[right][Train set size $60\%$ of the data.]{
    \includegraphics[width = 0.45\columnwidth]{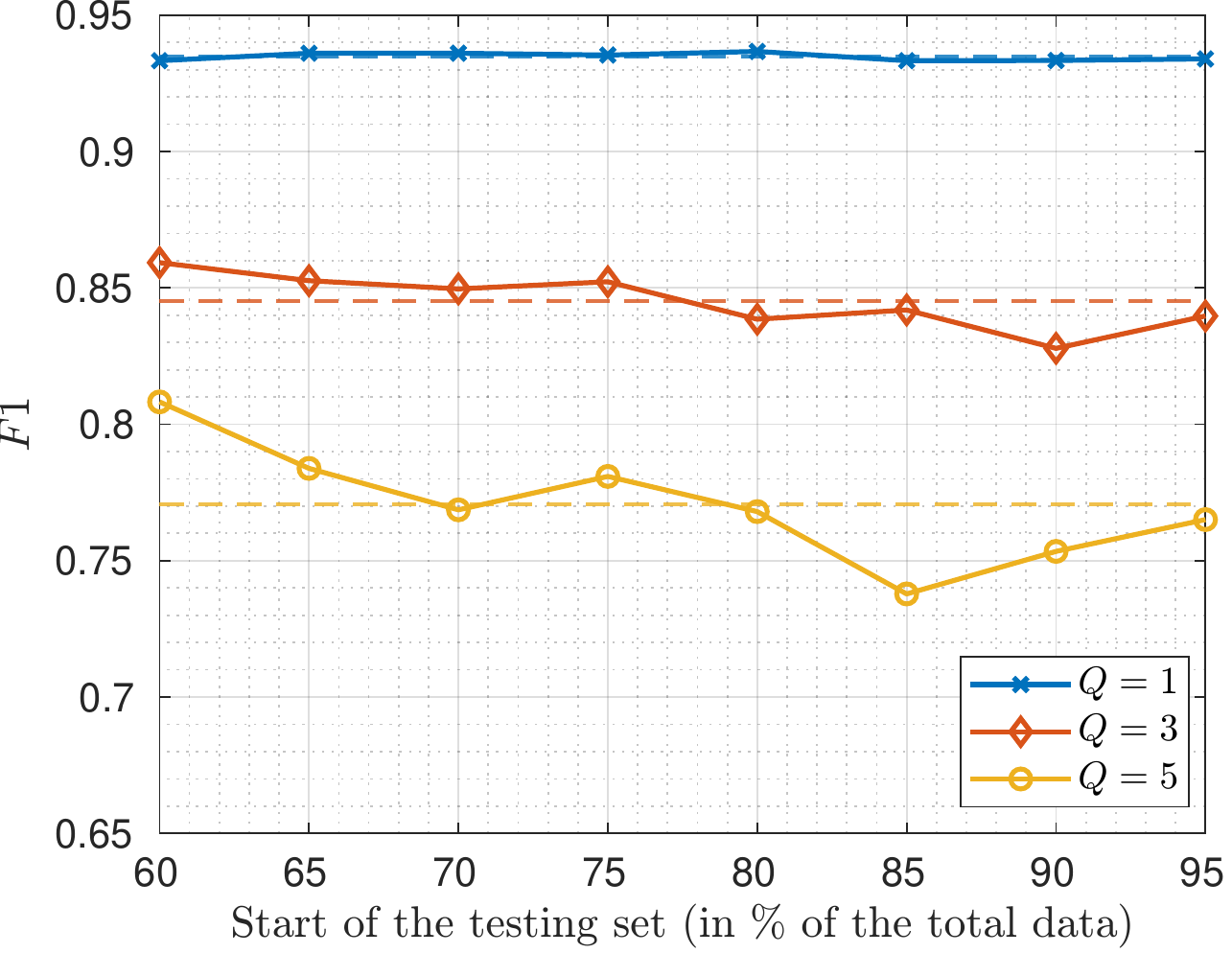}
    \label{fig:deg_6}}
 	\caption{$F1$ score for different ratios while varying the start of the testing set. Each values in the horizontal axes represent the score on a testing set composed of the $5\%$ of the data starting from the point forward. Dashed horizontal lines represent the mean of the $F1$ for each ratio.}
 	\label{fig:degradation}
\end{figure}

Initially, we train employing only the first $30\%$ of the data for each \ac{ev}. These results are shown in Figure~\ref{fig:deg_3} where it is possible to see a small difference in scores while shifting the testing set. However, we cannot blame the physical battery degradation for these results for a series of motivations. Firstly, the reduction is not really important in absolute terms (i.e., less than $0.1$ for $Q=1$) and can be due to different behaviors of the users in different periods (e.g., detaching the \ac{ev} before full \ac{soc}). Secondly, it is not the monotonous decrease that we would have expected. Instead, the scores seem not to follow any pattern, showing good classification results also in the last steps (i.e., testing set given by more recent \acp{ts}). Third, the small size of the training set employed can be a partial motivation for the random behavior of the scores.

In order to remove the train set size as a variable to create strange behavior in the results, we perform a second train using the $60\%$ of the data. By doing so, we obtain a chart with fewer points, as shown in Figure~\ref{fig:deg_6}. In this case, scores are almost constants with tiny variations which are expected in the context.  

This demonstrates how \emph{EVScout2.0} is able to perform well also considering a time distance of almost two years between the training and the testing data. Although not affected by the small battery degradation that could happen in this time frame, it could still be affected by unusual and unpredictable behaviors of the \acp{ev} owners or by the \ac{acn} scheduling algorithm, especially when considering highly unbalanced datasets and a small training set. Furthermore, this experiment can be seen as a remark on the need for a big training set as presented in Section~\ref{subsec:training_variation}. We will provide more analysis on the degradation as future work, considering datasets that span on more than two years.

\section{EVScout2.0 Performance: Comparison with EVScout}\label{sec:comparison}

In our previous work~\cite{brighente2021} we performed experiments similar to those discussed in previous sections but employing a different pipeline, a different feature extraction process, and a smaller dataset composed of only $22$ \acp{ev}. Since it was one of the first works on the privacy of the \ac{ev} charging systems, we use the results \cite{brighente2021} as a comparison baseline. In particular, we propose two different comparisons: one using EVScout on the new and extended dataset, and one using \emph{EVScout2.0} on the dataset employed in~\cite{brighente2021}. It is worth mentioning that we employed the same dataset used in~\cite{brighente2021} and not simply the same \acp{ev} updated with the new charges.  

\subsection{\textbf{EVScout on the new dataset}}
We tested EVScout~\cite{brighente2021} in our new dataset. To generate results comparable with those presented in this paper for \emph{EVScout2.0}, we employed the same dataset with $137$ valid \acp{ev}, even if \ac{ev} with less than eight charges could also be used in EVScout. We performed the same pre-processing phases, and we assess the performance for $N_{avg}=25$. Results are shown in Figure~\ref{fig:oldalg_newds} for different values of $Q$. We can see a clear dominance of \emph{EVScout2.0} with respect to its previous version in the new dataset composed by $137$ \acp{ev}. The difference in the classification performance ranges from about $0.15$ for $Q=5$ to more than $0.20$ for $Q=1$.

\subsection{\textbf{\emph{EVScout2.0} on the previous dataset}} 
\begin{figure}
    \subfigure[left][Comparison on the new dataset.]
    {\includegraphics[width = 0.45\columnwidth]{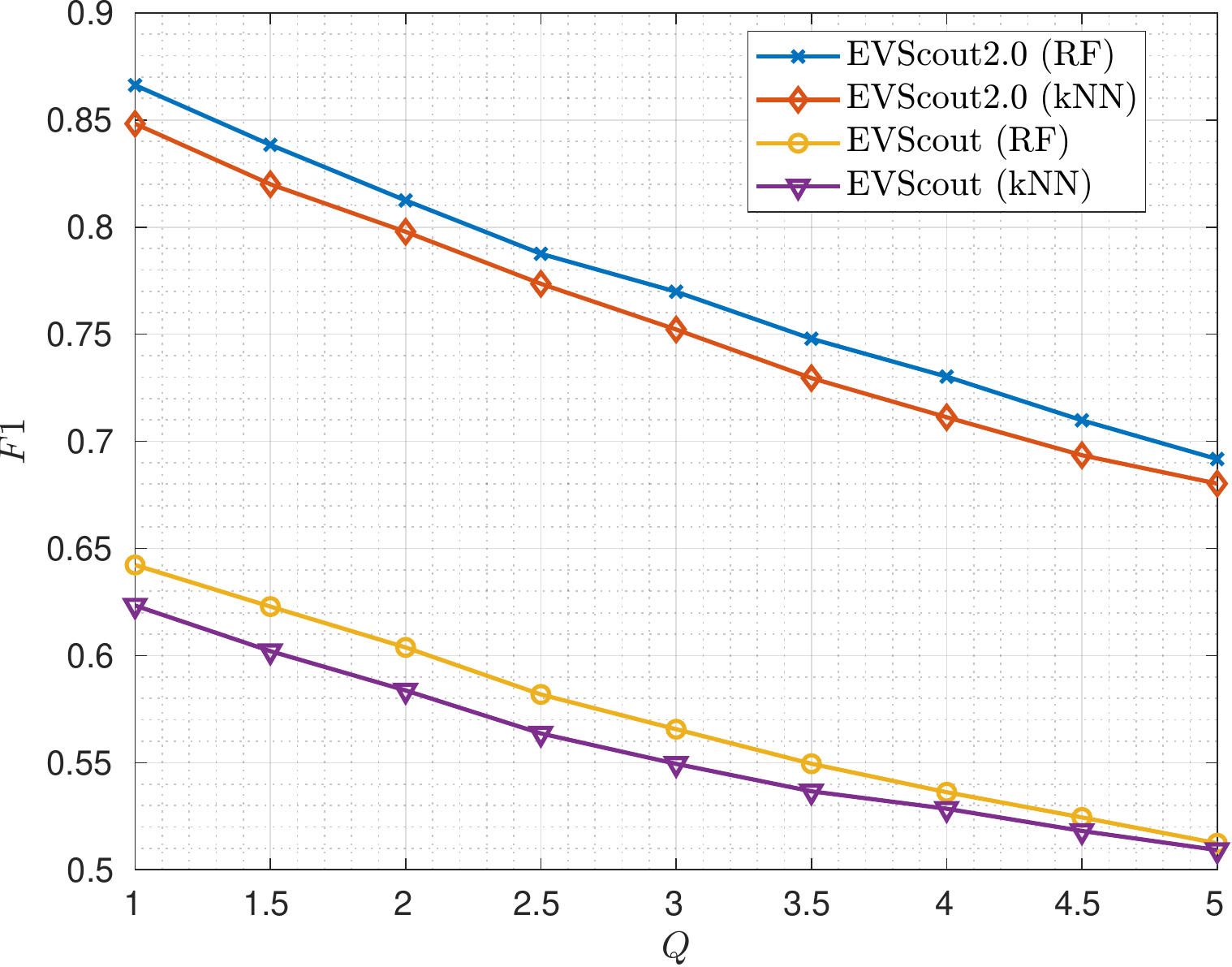}
    \label{fig:oldalg_newds}}
    \subfigure[right][Comparison on the original dataset.]{
    \includegraphics[width = 0.45\columnwidth]{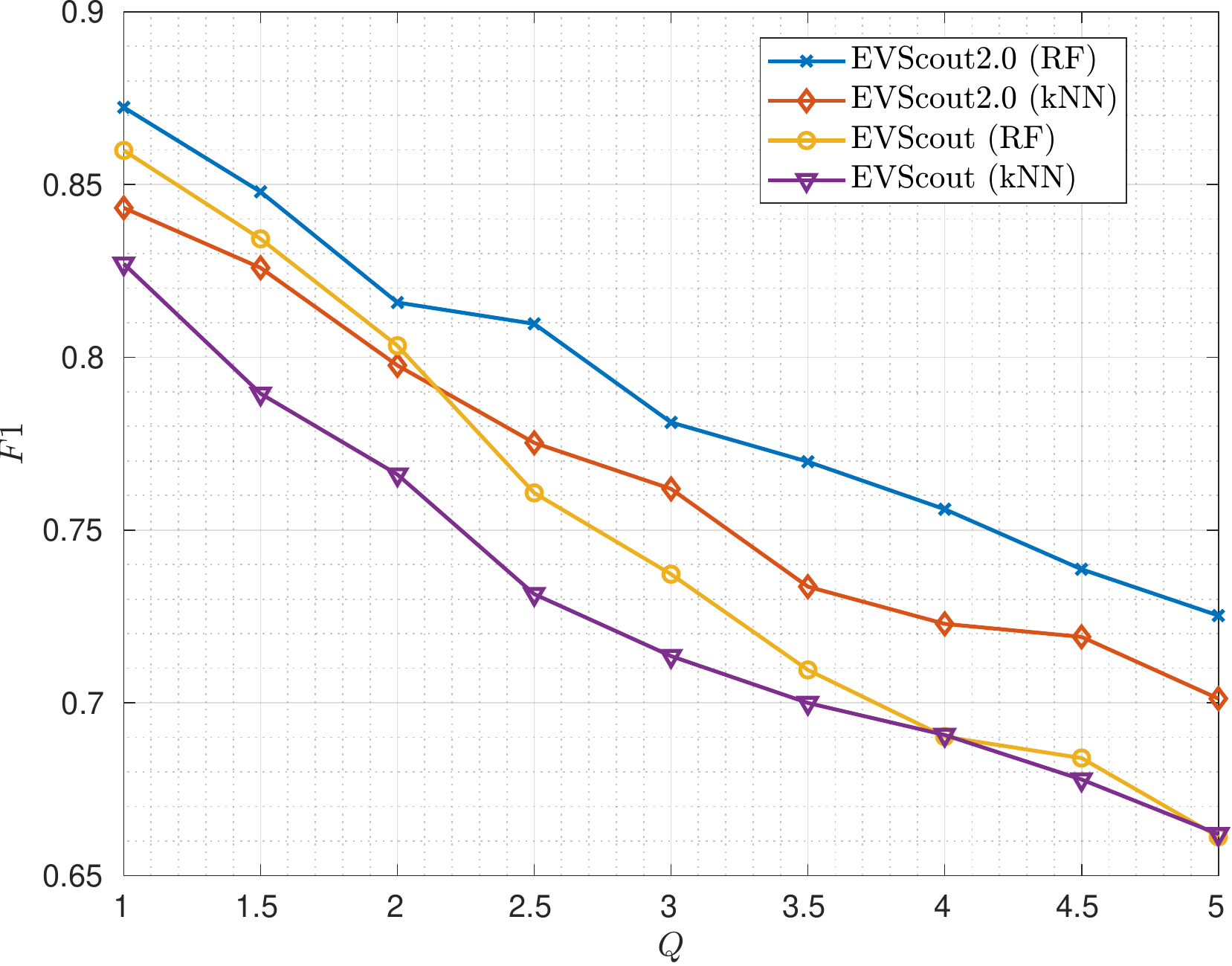}
    \label{fig:newalg_oldds}}
 	\caption{$F1$ values of the comparison with the previous work~\cite{brighente2021}.}
 	\label{fig:new_old}
\end{figure}

Furthermore, we test our new algorithm \emph{EVScout2.0} on the same dataset employed for the testing of EVScout. Since \emph{EVScout2.0} needs at least $8$ charging session comprising tails for each \ac{ev}, we have to discard four \acp{ev}, resulting in a total of $18$ \acp{ev}. We show the results of this experiment in Figure~\ref{fig:newalg_oldds}. Even if the enhancement is less pronounced with respect to the new dataset scenario, we can see a clear improvement in the performance of \emph{EVScout2.0} with respect to its previous version, especially considering high unbalancing $Q$. 

\section{Possible Countermeasures}\label{sec:countermeasures}

In order to deal with information leakage from smart meters, \cite{kalogridis2010privacy} proposes to obfuscate the communication between user and supplier through rechargeable batteries. This solution is effective in modifying the demand-response correlation in the measured data. This approach can be leveraged to mitigate the effects of \emph{EVScout2.0}, where the \ac{ev}'s battery drains current from a secondary battery which communicates with the \ac{evse}, masquerading the original battery's tail behavior. However, this approach incurs a higher implementation cost at both \ac{ev} and \ac{evse} sides, as the number of involved batteries doubles. Furthermore, the attacker may be able to extract features from the secondary battery and hence still be able to perform profiling.
A similar concept is exploited in \cite{shateri2020real}, where noise is added to smart meters data via an adversarial learning framework. This idea can be exploited to mitigate the effects of \emph{EVScout2.0}, by adding a suitable amount of noise to the current required by the \ac{ev}'s battery during the tail phase. However, this would imply redesigning how \acp{evse} manage the current required by \acp{ev}, as the added noise may mislead both the attacker and the \ac{evse}. The risk is, therefore, that the recharging process's efficiency drops. Other non-technical countermeasures are also possible. For instance, \ac{ev} owners can be educated to check the presence of suspicious devices attached to the \ac{evse}. Furthermore, running companies should often inspect their equipment for illegitimate devices and install closed-circuit TV to detect the presence of such devices.
Different solutions can hence be adopted to prevent user profiling. However, research should design proper countermeasures to ease the spreading of privacy-preserving \acp{ev} and \acp{evse}.

\section{Conclusions}\label{sec:conclusions}

\acp{ev} security is a novel topic that is rising many novel challenges in the protection of such systems. As we demonstrated, introducing a charging system that utilizes personal information can lead to privacy leakage and profiling attacks.
In this paper, we extended our previous work proposing \emph{EVScout2.0}. In particular, we extended the work in~\cite{brighente2021} by employing a bigger dataset, a novel feature extraction technique, and compared more algorithms for the classification task. We also show how real-world constraints such as limited training test size and battery degradation over time impact the classification quality.
With respect to the previous work, we employed a bigger dataset going from $22$ to $137$ \acp{ev}. We employed six different models capable of reaching precision and recall of $0.88$ for the unbalanced datasets, while still offering good results (precision $0.77$, recall $0.71$) with high unbalanced datasets. Furthermore, we evaluate the performance loss generated by a training set size reduction, showing how \emph{EVScout2.0} can reach good performances even with a small training set. In addition, we assess that the proposed algorithm is capable of correctly identifying \acp{ev} even if the model is trained with data two years early. Finally, we showed that \emph{EVScout2.0} is capable of attaining good classification performance, even in challenging scenarios such as highly imbalanced datasets or small training set, proving to be a viable and effective solution for \ac{ev} profiling.

\bibliographystyle{IEEEtran}
\bibliography{bibliography}
\end{document}